\documentclass[10pt,twocolumn]{IEEEtran}
\usepackage{lipsum}
\usepackage{balance}
\usepackage{soul,xcolor,lipsum,enumitem}
\usepackage[linesnumbered,ruled,vlined]{algorithm2e}

\SetCommentSty{mycommfont}
\usepackage{algorithmicx,comment,verbatim}
\usepackage{algpseudocode}
\usepackage{amsfonts}
\usepackage{amsmath}
\usepackage{upgreek}
\usepackage{amssymb}
\usepackage[ansinew]{inputenc} 
\usepackage{dsfont}
\usepackage{mathtools}
\usepackage{graphicx}
\usepackage[skip=1pt,font=scriptsize]{subcaption}
\usepackage{import}
\usepackage{multirow}
\usepackage{cite}
\usepackage[export]{adjustbox}
\usepackage{breqn}
\usepackage{mathrsfs}
\usepackage{acronym}
\usepackage{setspace}
\usepackage{stackengine}
\usepackage{steinmetz}
\usepackage{url}
\usepackage[skip=2pt,font=footnotesize]{caption}

\newcommand{\inprod}[2]{\langle #1, #2\rangle}

\usepackage{cancel}

\graphicspath{{./figures/}}
\newcommand{\secref}[1]{{Section~\ref{#1}}}
\newcommand\numberthis{\addtocounter{equation}{1}\tag{\theequation}}
\title{Mobility and Blockage-aware  Communications in Millimeter-Wave Vehicular Networks}

\author{Muddassar Hussain$^\dag$, Maria Scalabrin$^\ddag$, Michele Rossi$^\ddag$, and Nicol\`{o} Michelusi$^\dag$
\thanks{Part of this work appeared at IEEE ICC 2020~\cite{ICC2020}.}
\thanks{$^\ddag$School of Electrical and Computer Engineering, Purdue University, email: \{hussai13,michelus\}@purdue.edu}
\thanks{$^\dag$Department of Information Engineering, University of Padova, email: \{scalabri, rossi\}@dei.unipd.it}
\thanks{This research work has been supported, in part, by
NSF under grant \mbox{CNS-1642982}, by
 MIUR (Italian Ministry of Education, University and Research) through the initiative ``Departments of Excellence'' (Law 232/2016) and by the EU MSCA ITN project SCAVENGE ``Sustainable Cellular Networks Harvesting Ambient Energy'' (project no. 675891).}
 \thanks{Copyright (c) 2015 IEEE. Personal use of this material is permitted. However, permission to use this material for any other purposes must be obtained from the IEEE by sending a request to pubs-permissions@ieee.org}
 \vspace{-7mm}
}

\IEEEoverridecommandlockouts

\begin{document}
\setstcolor{red}
\setulcolor{red}
\setul{red}{2pt}

\maketitle

\begin{abstract}
Mobility may degrade the performance of next-generation vehicular networks operating at the millimeter-wave spectrum: frequent mis-alignment and blockages require repeated beam-training and handover, with enormous overhead.
Nevertheless, mobility induces temporal correlations in the communication beams and in blockage events. 
 In this paper, an adaptive design is proposed, that learns and exploits these temporal correlations to reduce the beam-training overhead and make handover decisions. 
  At each time-slot, the serving base station (BS) decides to perform either beam-training, data communication, or handover, under uncertainty in the system state. The decision problem is cast as a partially observable Markov decision process, 
with the goal to maximize the throughput delivered to the user, under an average power constraint. To address the high-dimensional optimization, an approximate  
 \emph{constrained point-based value iteration} (C-PBVI) method is developed,
 which simultaneously optimizes
 the primal and dual functions to meet the power constraint.
Numerical results demonstrate a good match between the analysis and a simulation
  based on 2D mobility and 3D analog beamforming via uniform planar arrays at both BSs and UE, and reveal that C-PBVI performs near-optimally, and outperforms a baseline scheme with periodic beam-training
by $38$\% in spectral efficiency. Motivated by the structure of C-PBVI, two heuristics are proposed, that trade complexity with sub-optimality, and achieve only 
 $4$\% and $15$\% loss in spectral efficiency.
Finally, the effect of mobility and multiple users on blockage dynamics is evaluated
 numerically, demonstrating superior performance  over the baseline scheme.
\end{abstract}
%
%
\vspace{-4mm}
\section{Introduction}
\label{sec:Introduction}
 Current sub-6GHz vehicular communication systems cannot support the demand of future applications such as autonomous driving, augmented reality and infotainment, due to limited bandwidth availability~\cite{choi2016millimeter}. To this end, new solutions are being explored that leverage the large bandwidth in the $30{-}300$GHz band, the so called \mbox{millimeter-wave} (\mbox{mm-wave}) spectrum. While communication at these frequencies is ideal to support high capacity demands, it relies on highly directional transmissions and is susceptible to blockages and \mbox{mis-alignment}, which are exacerbated in highly mobile and dense environments: the faster the environment mobile users operate in and the higher the density of users, the more frequent the loss of alignment and blockages, and the more resources need to be allocated to maintain beam alignment and perform handover to compensate for blockage.
   Mobility can thus be a source of severe overhead and performance degradation.
 Nevertheless, mobility induces temporal correlation in the communication beams and in blockage events.
   In this paper, we design adaptive strategies for beam-training, data transmission and handover, that exploit these temporal correlations to reduce the beam-training overhead and optimally trade-off throughput and power consumption.
   Our design allows to: 1) predict future beam-pointing directions and narrow down the beam search procedure
   to few likely beams, thus avoiding the enormous cost of exhaustive search; 2) more efficiently detect blockage and perform handover in response to it; 3) dynamically adjust the duration of the data communication phase based on predicted beam coherence times.
  However, two key questions arise: \emph{How do we leverage the system dynamics to optimize the communication performance? How much do we gain by doing so?} To address these questions, in this paper we envision the use of adaptive communication strategies and their formulation via partially observable (PO) Markov decision processes (MDPs) to optimize the decision-making process under uncertainty in the state of the system \cite{Pineau2006PointbasedAF}.

In the proposed scenario, two base stations (BSs) on both sides of a road link serve a user equipment (UE) moving along it. At any time, the UE is associated with one of the two BSs (the serving BS). To 
 enable directional data transmission (DT), the serving BS performs beam-training (BT); to compensate for blockage, it performs handover (HO) of the data traffic to the backup BS on the opposite side of the road link.
The goal is to design the BT/DT/HO strategy so as to maximize the throughput delivered to the UE, subject to an average power constraint.
 Mobility induces dynamics in the communication beams and in blockage events; we 
 show that these dynamics can be captured by a \emph{probabilistic state transition model}, which can be learned from interactions with the UE. However, the system state is not directly observable due to noise, beam imperfections, and detection errors; we thus
 formulate the optimization of the decision-making process as a
  constrained POMDP,
and develop an approximate \emph{constrained point-based value iteration} (C-PBVI) method to meet the average power constraint requirement: compared with PERSEUS~\cite{DBLP:journals/corr/abs-1109-2145}, originally proposed for unconstrained problems, 
C-PBVI allows to simultaneously optimize the primal and dual functions by decoupling the hyperplanes associated to reward and cost. We demonstrate its convergence numerically. 
Our numerical evaluations reveal a good match between the analysis based on a \emph{sectored antenna model} with Markovian state transitions, and a more realistic scenario with analog beamforming and \mbox{Gauss-Markov} mobility, hence demonstrating the effectiveness of our proposed scenario in more realistic settings:
simulations based on a 2D mobility model and 3D analog beamforming on both BSs and UE equipped with uniform planar arrays (UPA), demonstrate that \mbox{C-PBVI} performs near optimally,
  and outperforms a baseline scheme with periodic beam-training by up to $38$\% in spectral efficiency. Motivated by its structure, we design two heuristic policies with lower computational cost -- \mbox{belief-based} and \mbox{finite-state-machine-based} heuristics --
 and show numerically that they incur a small $4$\% and $15$\% degradation in spectral efficiency compared to \mbox{C-PBVI}, respectively. Finally, we demonstrate numerically 
 the effect of mobility and multiple users on the performance,
 based on the statistical blockage model developed in \cite{blockage_model}:
 the proposed low-complexity belief-based and finite-state-machine-based schemes achieve $50$\% and $25$\% higher spectral efficiency than the baseline scheme, respectively, demonstrating their robustness in mobile and dense user scenarios.

{\bf Related Work}: Beam-training design for \mbox{mm-wave} systems has been an area of extensive research in the past decade; various approaches have been proposed, such as beam sweeping~\cite{michelusi2018optimal}, estimation of angles of arrival (AoA) and of departure (AoD)~\cite{marzi}, and \mbox{data-assisted} schemes~\cite{inverse_finger}. Despite their simplicity, the overhead of these algorithms may offset the benefits of beamforming in highly mobile environments~\cite{choi2016millimeter}. While wider beams require less beam-training, they result in a lower beamforming gain, hence a smaller achievable capacity~\cite{7744807}. 
Contextual information, such as GPS readings of vehicles~\cite{inverse_finger}, may alleviate this overhead, but it does not eliminate the need for beam-training due to noise and inaccuracies in GPS acquisition. Thus, the design of schemes that alleviate the beam-training overhead is of great importance. 
 
In most of the aforementioned works, a priori information on the vehicle's mobility as well as blockage dynamics is not leveraged in the design of communication protocols. In contrast, \emph{we contend and demonstrate numerically that learning and exploiting such information via adaptive communications can greatly improve the performance of \mbox{mm-wave} networks}~\cite{va2016beam}. 
In our previous work~\cite{michelusi2018optimal}, we bridged this gap by leveraging worst-case mobility information to design beam-sweeping and data communication schemes; in~\cite{scalabrin2018beam}, we designed adaptive strategies for BT/DT that leverage a Markovian mobility model via POMDPs, but with no consideration of blockage (hence no handover).
A distinctive feature of the \mbox{mm-wave} channel is its highly dynamic link quality, due to the occurrence of blockages on very short time-scales~\cite{mezzavilla2016mdp}. In this respect, handover represents a fundamental functionality to preserve communication in the event of link obstruction; however, it is challenging to implement it in \mbox{mm-wave} networks, since the mm-wave link quality needs to be accurately tracked and blockages need to be quickly detected -- a difficult task to accomplish using highly directional communications. Therefore, \mbox{MDP-based} handoff strategies proposed for \mbox{sub-5GHz} systems cannot be readily applied~\cite{pan2012mdp,stevens2008mdp}.
 In this paper, we develop \mbox{feedback-based} techniques to quickly detect blockages, and enable a \mbox{fully-automatic} and data-driven optimization of the handover strategy via POMDPs.

Recent work~\cite{alkhateeb2018deep,va2018online,second-best,javdi,TWC2019} that applies machine learning to \mbox{mm-wave} networks reveal a growing interest in the design of schemes that exploit side information to enhance the overall network performance. 
For example,~\cite{alkhateeb2018deep} develops a coordinated beamforming technique using a combination of deep learning and \mbox{ray-tracing}, and demonstrates its ability to  efficiently adapt to changing environments. More recent solutions are based on \mbox{multi-armed} bandit, by leveraging \emph{contextual information} to reduce the training overhead as in~\cite{va2018online}, or the beam alignment feedback to improve the beam search as in~\cite{second-best,javdi,TWC2019}.
    However, no handover strategies are considered in these works, resulting in limited ability to combat blockage.
    In addition, these works neglect the impact of realistic mobility and blockage processes on the performance.
Compared to this line of works, in this paper we design adaptive communication strategies that leverage learned statistical information on the mobility and blockage processes in the selection of BT/DT/HO actions, with the goal to optimize the average \mbox{long-term} communication performance of the system. Our proposed approach is in contrast to strategies that
either use non-adaptive algorithms \cite{alkhateeb2018deep}, lack a handover mechanism \cite{va2018online,second-best,javdi,TWC2019}, or assume a non realistic mobility pattern in their design.\\
\label{contributions}
{\bf Our Contributions}: 
\begin{itemize}[leftmargin=*]
\item We define a POMDP framework  to optimize the BT/DT/HO strategy in a \mbox{mm-wave} vehicular network, subject to 2D mobility of the UE and time-varying blockage, with the goal to maximize throughput subject to an average power constraint;
 \item
 We propose a novel feedback mechanism for BT, which reports the ID of the strongest \mbox{BS-UE} beam pair if the received power is above a threshold (a design parameter), otherwise it reports $\emptyset$ to indicate \mbox{mis-alignment} or blockage. We analyze its detection performance in closed form; 
\item 
To address the complexity of POMDPs, we design \mbox{C-PBVI}, a constrained \mbox{point-based} value iteration method.
In order to incorporate the average power constraint, we extend PERSEUS~\cite{DBLP:journals/corr/abs-1109-2145}, originally designed for unconstrained POMDPs, via a Lagrangian formulation, the separation of hyperplanes for \mbox{reward-to-go} and \mbox{cost-to-go} functions, and a dual optimization step to solve the constrained problem. We demonstrate its convergence numerically; 
\item Inspired by the \mbox{C-PBVI} policy, we propose two heuristic schemes that trade complexity with sub-optimality,
  namely belief-based (B-HEU) and finite-state-machine-based (FSM-HEU) heuristic policies. We analyze the performance of FSM-HEU in closed form.  
\end{itemize}
\indent The rest of the paper is organized as follows. In~\secref{sec:System_Model}, we introduce the system model: signal and channel models (\secref{sec:sig_model}), codebook structure (\secref{sec:codebook}),
mobility and blockage dynamics (\secref{sec:beam_blockage_dynamics}),  sectored antenna model (\secref{sec:sect_antenna}), and BT/DT mechanisms (\secref{sec:BT_DT}). We provide the POMDP formulation in  \secref{sec:POMDP} and its optimization via \mbox{C-PBVI} in~\secref{sec:Problem_Optimization}. In \secref{sec:heuristic}, we present the two heuristic policies \mbox{B-HEU} and \mbox{FSM-HEU}, along with a mathematical analysis of the latter. Selected  numerical results are presented in~\secref{sec:numres}, while~\secref{sec:Conclusions} reports some concluding remarks. 
\vspace{-4mm}
\section{System Model}
\label{sec:System_Model}
\begin{figure}[t!]	
	\centering
	\includegraphics[trim = 0 0 0 10,clip,width=0.9\columnwidth]{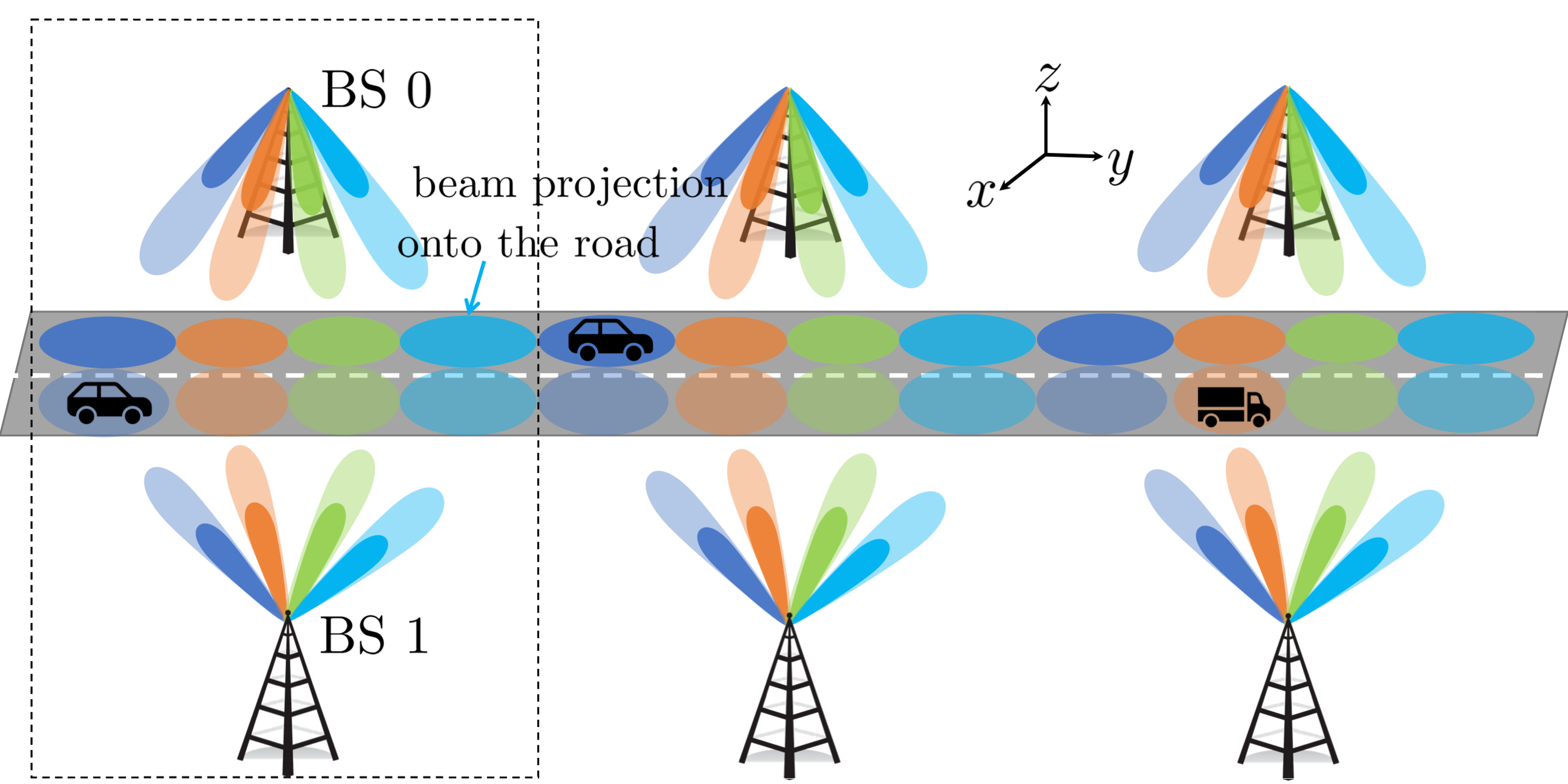}
	\caption{A cell deployment with BSs on both side of the road.}\label{figure:Fig_scenario}
\end{figure}

We consider the scenario of Fig.~\ref{figure:Fig_scenario}, where
multiple base stations (BSs) serve user equipments (UEs) moving along a road. At any time, each UE is associated with one BS -- the \emph{serving BS}.
Each UE and the serving BS use beamforming with large antenna arrays to achieve 
directional data transmission (DT);  they use beam-training (BT) to maintain alignment.
The communication links are subject to time-varying blockages, which cause the signal quality to drop abruptly and DT to fail. 
As soon as the serving BS detects blockage, it may decide to perform handover (HO) to the BS on the other side of the road, which then continues the process of BT/DT/HO, until either another blockage event is detected, or the UE exits the coverage area of the two BSs.
 
 \par 
 In this work, we focus on a specific segment of the road link covered by a pair of BSs and a single UE,\footnote{\label{fn:multiuser}The proposed system model and  techniques can be applied to a multi-user scenario by partitioning the BS resources using orthogonal frequency division multiple access (OFDMA) and multiple RF chains or time division duplexing (TDD)\cite{oma}.} as depicted in the framed area of Fig.~\ref{figure:Fig_scenario}.
Within this segment, the BT/DT/HO  process continues until the UE exits the coverage region of the two BSs, denoted by the area $\mathcal X\subset\mathbb R^2$.
 In this context, we investigate the design of the BT/DT/HO strategy during a transmission episode, defined as the time interval between the two instants when the UE enters and exits the coverage area of the two BSs. The goal is to maximize the average throughput delivered to the UE subject to an average power constraint.
Note that, when the  episode terminates,  the UE enters the coverage area of another pair of BSs, and the same analysis may be applied to each segment traversed.

Time is discretized into \mbox{time-slots} of duration $\Delta_{\rm t}$, 
corresponding to the transmission of  a beacon signal during BT or of a data fragment during DT.  
 Next, we describe the signal, channel and UE mobility and blockage dynamics models used throughout the paper.
\vspace{-5mm}
\subsection{Signal and Channel Models} 
\label{sec:sig_model}
Let $I\in\{0,1\} \triangleq \mathcal I$ denote the index of the serving BS at time $k$.
 Let $\mathbf x_k{\in}\mathbb C^L$ be the transmitted signal  with $\mathbb E[\Vert \mathbf x_k\Vert_2^2]{=}L$, where $L$ denotes the number of symbols transmitted. 
The received signal at the UE is expressed as 
\begin{align}
\label{eq:signal_model}
\mathbf y_k = \sqrt{P_k} \mathbf f_k^H \mathbf{H}_k^{(I)} \mathbf{c}_{k}\mathbf x_k +\mathbf w_k,
\end{align}
where $P_k$ is the average transmit power of the serving BS $I$; 
 $\mathbf{c}_{k} {\in} \mathbb{C}^{M_{\rm tx}^{(I)}\times 1}$ and $\mathbf{f}_{k} {\in} \mathbb{C}^{M_{\rm rx}\times 1}$  are unit-norm  beamforming vectors 
 with 
$M_{\rm tx}^{(I)}$ and  $M_{\rm rx}$ antenna elements at BS $I$ and the reference UE, respectively;   $\mathbf{H}_k^{(I)} {\in} \mathbb{C}^{M_{\rm rx}\times M_{\rm tx}^{(I)}}$ is the channel matrix; $\mathbf w_k {\sim} \mathcal{CN}(0, \sigma_w^2\mathbf I)$ with $\sigma_w^2{=}(1+F)N_0W_{\rm tot}$ is
additive white Gaussian noise, $N_0$ is the noise power spectral density, $W_{\rm tot}$ is the signal bandwidth, $F$ is the receiver noise figure.\\
\indent In this paper, we model $\mathbf{H}_k^{(I)}$ as a single line of sight (LOS) path with binary blockage \cite{blockage}
and diffuse multipath \cite{diffuesedTSP},
\begin{align*}
\mathbf{H}_k^{(I)}{=}&\underbrace{\sqrt{M_{\rm tx}^{(I)}M_{\rm rx}}B_k^{(I)} h_k^{(I)} \mathbf{d}_{\rm rx}(\theta^{(I)}(X_k)) \mathbf{d}_{\rm tx}^{(I)}(\phi^{(I)}(X_k))^H}_{\mathbf H_{k,{\rm LOS}}^{(I)}}  \\
&+\underbrace{ \sum_{l=1}^{N_{\rm DIF}}\!\!\sqrt{M_{\rm tx}^{(I)}M_{\rm rx}} \tilde h_{k,l}^{(I)}\mathbf{d}_{\rm rx}(\tilde\theta_{k,l}^{(I)}) \mathbf{d}_{\rm tx}^{(I)}(\tilde\phi_{k,l}^{(I)})^H}_{\mathbf H_{k,{\rm DIF}}^{(I)}}
\!\!,
\end{align*} 
\vspace{-3mm}\\
where $B_k^{(I)}{\in}\{0,1\}$ denotes the binary blockage variable of BS $I$, equal to $1$ if the LOS path is unobstructed, equal to 0 otherwise;
 $\mathbf{d}_{\rm tx }^{(I)}(\phi){\in}\mathbb{C}^{M_{\rm tx}^{(I)}}$ and  $\mathbf{d}_{\rm rx}(\theta) \in \mathbb{C}^{M_{\rm rx}}$  are the unit-norm array response vectors of  BS $I$ and UE, as a function of the AoD $\phi$ and AoA $\theta$ (note that these include both azimuth and elevation information for UPAs); $\phi^{(I)}(X_k)$ and $\theta^{(I)}(X_k)$ are the AoD and AoA of the LOS path with respect to BS $I$ and the UE in position $X_k \in \mathcal X$;\footnote{Note that the AoA  $\theta^{(I)}(X_k)$  should also depend on the angle of rotation  (azimuth and elevation) of the antenna array of the UE; herein, we assume that it only depends on the UE position $X_k$. This is a good approximation in vehicular networks, where the antenna array may be mounted on the rooftop of the vehicle; 
 the more general case with non-fixed array orientation can be addressed by including 
 the angle of rotation information in the AoA, which may be estimated using a gyroscope sensor \cite{Shim_2014}.} $h_k^{(I)} {\sim} \mathcal{CN}(0,\sigma_{h,I}^2)$ is the complex channel gain of the LOS component, i.i.d. over slots,
 with $\sigma_{h,I}^2{=}1/\ell(d_I(X_k))$; $\ell(d_I(X_k))=(4\pi d_I(X_k)/\lambda_c)^2$ denotes the pathloss as function of the BS $I$-UE distance $d_I(X_k)$;
$\lambda_c{=}c/f_c$ is the wavelength.
Finally, $\mathbf H_{k,{\rm DIF}}^{(I)}$ denotes the channel corresponding to diffuse multipath components with coefficients $\tilde h_{k,l}$,  AoD $\tilde \phi_{k,l}^{(I)}$ and AoA $\tilde \theta_{k,l}^{(I)}$; we model $\mathbf H_{k,{\rm DIF}}^{(I)}$ as \mbox{zero-mean} complex Gaussian, with i.i.d. entries (over time and over antennas), each with variance $\sigma_{{\rm DIF},I}^2$.
 These components have been shown to be much weaker than the LOS path (up to 100$\times$ weaker at a \mbox{BS-UE} distance of only 10 meters  \cite{blockage}), so that $\sigma_{{\rm DIF},I}^2{\ll}\sigma_{h,I}^2$.
\indent Then, 
letting $G_{{\rm tx}}^{(I)}(\mathbf{c}_k,x){=}M_{\rm tx}^{(I)}|\mathbf{d}_{\rm tx}^{(I)}(\phi^{(I)}(x))^H\mathbf{c}_k|^2$ and $G_{{\rm rx}}(\mathbf{f}_k,x){=}M_{\rm rx}|\mathbf{d}_{\rm rx}(\theta^{I}(x))^H\mathbf{f}_k|^2$
be the beamforming gains of the serving BS $I$ and UE, respectively, with respect to the LOS path, and $\Theta_{k}{=}\angle{\mathbf{d}_{\rm tx}^{(I)}(\phi^{(I)}(X_k))^H\mathbf{c}_k} + \angle{\mathbf{f}_k^H\mathbf{d}_{\rm rx}(\theta^{(I)}(X_k))} $ be the unknown phase of the overall gain, the signal received at the UE can be expressed as 
\begin{align*}
\label{sigmodel}
\!\mathbf y_k{=} &\sqrt{P_k} \!\left[\!B_k^{(I)}h_k^{(I)} \sqrt{G_{{\rm tx}}^{(I)}(\mathbf{c}_k,X_k) G_{{\rm rx}}(\mathbf{f}_k,X_k)} e^{j\Theta_{k}}{+} \Omega_{k}^{(I)} \!\right]\!\mathbf x_k \\
& \qquad\qquad+ \mathbf w_k,\numberthis
\end{align*}
where $\Omega_{k}^{(I)} \triangleq \mathbf f_k^H \mathbf H_{k,{\rm DIF}}^{(I)}\mathbf c_k\sim  \mathcal{CN}(0,\sigma_{{\rm DIF},I}^2)$ is the contribution due to the diffuse multipath channel components.
The SNR averaged over the fading coefficients is then given as 
 \begin{align*}
 \label{eq:SNR}
\! {\rm SNR}_k{= }
 \frac{P_k}{\sigma_w^2}
\left[ B_k^{(I)}\frac{G_{{\rm tx}}^{(I)}(\mathbf{c}_k,X_k) G_{{\rm rx}}(\mathbf{f}_k,X_k)}{\ell(d_I(X_k))}+\sigma_{{\rm DIF},I}^2\right].
\numberthis
 \end{align*}
\vspace{-7mm}
\subsection{Codebook Structure}
\label{sec:codebook}
 Each BS has a codebook of beamformers to cover the intended coverage region $\mathcal X$ on the road. The beamforming codebook of BS $I$ is denoted by $\mathcal C_{I} {\triangleq} \{\mathbf c_{I,1},\ldots, \mathbf c_{I,|\mathcal C_I|} \}$. The UE uses the codebook $\mathcal F {\triangleq} \{\mathbf f_1,\ldots, \mathbf f_{|\mathcal F|} \}$. Let $\mathcal V_I{\triangleq}\mathcal C_I \times \mathcal F$ denote the joint codebook containing all possible beamforming codeword pairs of BS $I$ and UE. We index these codeword pairs by the \emph{beam pair index} (BPI), with values in $\bar {\mathcal S}_I \triangleq \{1,2,\ldots,|\mathcal C_I||\mathcal F|\}$; let $(\mathbf c_I^{(j)},\mathbf f_I^{(j)})$ be the $j$th such pair, with
 $j{\in}\bar {\mathcal S}_I$.
  With this definition, note that, if the UE is in position $X_k{=}x$ and is being served by BS $I$, then the maximum beamforming gain is achieved with the strongest BPI (SBPI),  which also yields the maximum SNR in \eqref{eq:SNR}, defined as 
\begin{align}
\label{sbpi}
 s^*_I(x) \triangleq \arg \max_{j \in \bar {\mathcal S}_I} G_{{\rm tx}}^{(I)}(\mathbf{c}_I^{(j)},x) G_{{\rm rx}}(\mathbf{f}_I^{(j)},x).
 \end{align}
\vspace{-4mm}\\
Let $\mathcal S_I \triangleq \left\{ s^*_I(x): x \in \mathcal X \right\} \subseteq \bar {\mathcal S}_I$  be the set of SBPIs across all possible UE positions.
 Note that this set can be constructed over time utilizing the feedback from the UE and excluding the BPIs that do not yield significant signal power \cite{inverse_finger}. It follows that the directional communication between BS $I$ and UE can be achieved by restricting the choice of beamforming codewords to the optimal set $\mathcal S_I$, since any other beam pair achieves lower SNR. This can be obtained using a coordinated beamforming strategy where, before start of BT or DT, the serving BS $I$ and UE coordinate to select a subset of BPIs from the set $\mathcal S_I$ to be scanned synchronously during BT or used for DT, as explained in Section~\ref{sec:BT_DT}.
\vspace{-3mm}
\subsection{Mobility and Blockage Dynamics}
\label{sec:beam_blockage_dynamics}
Note that, to achieve directional communication, the pair of BS $I$ and UE should
detect the SBPI $s^*_I(X_k)$ via beam-training -- a source of severe overhead; the mobility of the UE along the road induces temporally correlated dynamics on the SBPI $s^*_I(X_k)$, which may be exploited to reduce the training overhead via POMDPs. Similarly, the blockage state exhibits temporal and spatial correlations, which can be exploited to efficiently detect/predict blockages and perform HO if needed. To define such POMDP model, we now define 
a Markov model on the SBPI and blockage states, induced by the UE mobility.
Let $S_k=(s^*_0(X_k),s^*_1(X_k))$ be the pair of SBPIs at both BSs, 
taking values from  $ \mathcal S \triangleq \{\left(s^*_0(x),s^*_1(x)\right): x \in \mathcal X \} $.
 Let $B_k \triangleq (B_k^{(0)},B_k^{(1)}) \in \{0,1 \}^2$ be the pair of binary blockage states with $B_k^{(I)}$ denoting the blockage with respect to BS $I$. Then, the one-step transition probability of $(S_k,B_k)$  is expressed as
\begin{align}
\label{eq:trans_probs_MM}
&\mathbf P_{s'b'|sb}\triangleq\mathbb P (S_{k+1} = s',B_{k+1} = b'|S_k=s, B_k = b)\\&
\!\!\!{=}
\underbrace{\mathbb P(S_{k+1} {=} s'|S_k{=}s)}_{{\mathbf S_{s'|s}}} \underbrace{\mathbb P(B_{k+1} {=} b'|B_k {=} b,S_k{=}s, S_{k+1} {=} s')}_{{\mathbf B_{b'|bss'}}}.
\nonumber
\end{align}
Here, it is assumed that the next SBPI $S_{k+1}$ is independent of the current blockage state $B_k$, given the current beam index pair $S_k$ (indeed, the dynamics of SBPI depend solely on UE mobility). Note that $\sum_{s',b'}\mathbf P_{s'b'|sb}\leq 1$, since the UE might exit the coverage area of the two BSs. In practice, \eqref{eq:trans_probs_MM} can be estimated based on estimated time-series of SBPI and blockage pairs, $\{(\hat s_k,\hat b_k,\hat s_{k+1},\hat b_{k+1}), k\in T_{\rm sound}\}$, which in turn may be acquired at times $k\in T_{\rm sound}$ via exhaustive search beam-training methods. Based on these \mbox{time-series}, the BSs can estimate the transition probabilities in \eqref{eq:trans_probs_MM} as
\begin{align}
\label{eq:mob_est}
&\!\!\!\hat {\mathbf S}_{s'|s} = \frac{\sum_{k\in T_{\rm sound}} \chi(\hat s_k = s, \hat s_{k+1} = s')}{\sum_{k\in T_{\rm sound}} \chi(\hat s_k = s)},
\end{align}
\begin{align}
\label{eq:block_est}
&\!\!\!{\hat{\mathbf B}_{b'|bss'}}
{=} \frac{\sum_{k\in T_{\rm sound}} \chi(\hat s_k {=} s, \hat B_k {=} b, \hat s_{k+1} {=} s',\hat B_{k+1} {=}b')}{\sum_{k\in T_{\rm sound}} \chi(\hat s_k {=} s, \hat B_k {=} b, \hat s_{k+1} {=} s')},\!
\end{align}
where $\chi(\cdot)$ is the indicator function.
\label{discussion:Estimation}Note that the estimates $\hat {\mathbf S}_{s'|s}$ and $\hat{\mathbf B}_{b'|bss'}$ can be improved over time as more samples of $(\hat s_k,\hat b_k,\hat s_{k+1},\hat b_{k+1})$ become available.
This approach does not require a dedicated learning phase; instead, estimated \mbox{time-series} can be collected based on 
beam-training and data communication feedback, so that the estimation overhead is minimal. Following their updates, the proposed policies can be updated accordingly. As more and more samples are collected, the estimation accuracy improves, leading to policies that more optimally leverage the mobility and blockage dynamics within the environment, yielding a more efficient use of resources.

\vspace{-6mm}                                                                                                                                                                                                                                                                                                                                                                                                                                                                                                                                                                                                                                                                                                                                                                                                                                                                                                                                                                                                       \subsection{Sectored antenna model} 
\label{sec:sect_antenna}                         
In this paper, we use the \emph{sectored antenna model} to approximate the  beamforming gain, as also used in~\cite{va2016beam,TWC2019}. 
As we will show in \secref{sec:numres}, when coupled with an appropriate design of the BSs beamforming codebooks $\mathcal C_I, I \in \mathcal I$ and of the UE beamforming codebook $\mathcal F$~\cite{noh}, the sectored model provides an accurate and analytically tractable approximation of the actual beamforming gain.
Consider the BPI $j\in \mathcal S_I$ and let $G^{(I)}(j,x) \triangleq G_{\rm tx}^{(I)}(\mathbf c_I^{(j)},x) G_{\rm rx}^{(I)}(\mathbf f_I^{(j)},x)$ be the overall gain between BS $I$ and UE position $x$, under the beamforming codeword pair $(\mathbf c_I^{(j)},\mathbf f_I^{(j)})$.
 Under the sectored model, if the UE is aligned with BS $I$ under the BPI $j$, i.e., its position $x$ is such that the SBPI $s_I^*(x)=j$, then the \emph{aligned gain} satisfies $G^{(I)}(j,x){\gg}1$ with \mbox{gain-to-pathloss} ratio $G^{(I)}(j,x)/\ell(d^{(I)}(x)) \approx \Upsilon_I^{(j)},{\forall x:} j{=}s_I^*(x)$. On the other hand, if the UE is mis-aligned with BS $I$ under the BPI $j$, i.e., 
$s_I^*(x)\neq j$, then the mis-aligned beamforming gain of BPI $j\in \mathcal S_I$ is such that $G^{(I)}(j,x) {\approx} g_{j}^{(I)} {\ll}1,{\forall x:} j{\neq}s_I^*(x)$ (i.e., it is small and equal to the sidelobe gain $g_{j}^{(I)} $ for all positions $x$ such that $j$ is not the SBPI).
Based on this model, we now derive expressions for the transmission power to achieve a target SNR at the receiver. We denote the case with the aligned beam pair and no blockage ($j=s_I^*(x)$ and $b_I{=}1$) as ``active SBPI'' and the complementary case of blockage or UE in the sidelobe ($j\neq s_I^*(x)$ or $b_I{=}0$) as ``inactive SBPI''. In the case of active SBPI, from \eqref{eq:SNR} we have
\begin{align}
\label{eq:PSNR}
{\rm SNR}_{\rm act}
{=}\frac{ P_j^{(I)}}{\sigma_w^2}\left[\Upsilon_j^{(I)}{+}\sigma_{{\rm DIF},I}^2\right]
 {\Leftrightarrow} 
 P_j^{(I)}{=}\frac{\sigma_w^2{\rm SNR}_{\rm act}}{\Upsilon_j^{(I)}{+}\sigma_{{\rm DIF},I}^2},
\end{align}
which yields the transmission power to achieve a target SNR equal to ${\rm SNR}_{\rm act}$ in case of active SBPI. In the case of  inactive SBPI, we can express the SNR in \eqref{eq:SNR} using \eqref{eq:PSNR} as 
\begin{align*}
{\rm SNR}_{\rm iact}
 &=  \frac{P_j^{(I)}}{\sigma_w^2} \left[ {B^{(I)}}\frac{G^{(I)}(j,x)}{\ell(d_I(x))} {+} \sigma_{{\rm DIF},I}^2\right]\\
 & =
\left[B^{(I)} 
 \frac{G^{(I)}(j,x)}{\ell(d_I(x))} {+}  \sigma_{{\rm DIF},I}^2\right]\frac{{\rm SNR}_{\rm act}}{ \Upsilon_j^{(I)}+ \sigma_{{\rm DIF},I}^2}.\numberthis
\label{eq:SLSNR}
\end{align*}
Note that, to help the BS detect the  inactive SBPI condition, this value of SNR should be as small as possible; for this reason, we determine the worst case SNR under inactive SBPI by maximizing \eqref{eq:SLSNR} over all possible blockage states
${B^{(I)}}{\in}\{0,1\}$, \mbox{mis-aligned} beam $j$ and UE position $x {\in} \mathcal X$, as
\begin{align*}
\label{worstcase}
&{\rm SNR}_{\rm iact} \\
 &{\leq}
 \max_{x\in \mathcal X}\max_{j{\in}\mathcal S_I \setminus\{s_I^*(x) \}}
  \left[B^{(I)} 
 \frac{G^{(I)}(j,x)}{\ell(d_I(x))} {+}  \sigma_{{\rm DIF},I}^2\right]\frac{{\rm SNR}_{\rm act}}{ \Upsilon_j^{(I)}{+} \sigma_{{\rm DIF},I}^2}
 \\& \triangleq \rho_I {\rm SNR}_{\rm act}.
 \numberthis
\end{align*}
In other words, to achieve a target ${\rm SNR}_{\rm act}$ within the mainlobe, the BS should transmit with power given by \eqref{eq:PSNR}; 
however, if the signal is blocked or the UE receives on the sidelobe (or both), the  associated  worst-case SNR  is  $\rho_I {\rm SNR}_{\rm act}$.\footnote{For the sake of analytical tractability, $\rho_I$ (found by maximizing over $j\neq s_I^*(x)$) is
the worst case over the BPI $j \in \mathcal S_I$. 
The model can be generalized to express the dependence of $\rho_I$ on $j$, 
leading to a more complicated BT feedback analysis, possibly not in closed form. }
In this case, data transmission is in outage since $\rho_I\ll 1$  (numerically, we found $\rho_I= -15$dB, $\forall I$ based on the setup of \secref{sec:numres}).
 
\vspace{-3mm}
\subsection{Beam-Training (BT) and Data Transmission (DT)} 
\label{sec:BT_DT}

We now introduce the BT and DT operations.

\textbf{BT phase:} At the start of a BT phase, the serving BS $I$ selects a set of BPIs ${\mathcal S}_{\rm BT} {\subseteq} \mathcal S_I$ over which the beacons $\mathbf x_k$ are sent,
and a target SNR ${\rm SNR}_{\rm BT}$.
The beacon transmission is done in sequence over $|{\mathcal S}_{\rm BT}|$ time-slots, using one slot for each BPI $j{\in} {\mathcal S}_{\rm BT}$, with the serving BS transmitting using the beamforming vector $\mathbf c_I^{(j)}$, and the UE synchronously receiving using the combining vector $\mathbf f_I^{(j)}$. Therefore, the duration of the BT phase is $T_{\rm BT}{\triangleq}|{\mathcal S}_{\rm BT}|{+}1$, including the last slot for feedback signaling from the UE to the BS.
 Let $i\in\{0,{\ldots},T_{\rm BT}-2\}$ be the $i$th time-slot of the BT phase, and $j_i\in{\mathcal S}_{\rm BT}$ be the BPI scanned by the BS $I$ and UE in this slot. The UE processes the received signal $\mathbf{y}_{k+i}$ with a matched filter,
 \begin{align}
 \Gamma_{j_i}\triangleq\frac{|\mathbf{x}_{k+i}^H \mathbf{y}_{k+i}|^2}{(1+F)N_0 W_{\rm tot} \Vert\mathbf{x}_{k+i}\Vert_2^2}.
 \label{zsi}
\end{align}
 Upon collecting the sequence $\{\Gamma_{j},\forall j\in{\mathcal S}_{\rm BT}\}$,
the UE generates the feedback signal
\begin{align}
\label{btfb}
Y = \begin{cases}
{j}^*\triangleq \arg\max_{{j}\in{\mathcal S}_{\rm BT}}\Gamma_{j},& \max_{{j}\in{\mathcal S}_{\rm BT}} {\Gamma_{j}}>\eta_{\rm BT}^{(I)},\\
\emptyset,& \max_{{j}\in{\mathcal S}_{\rm BT}} {\Gamma_{j}}\leq\eta_{\rm BT}^{(I)}.
\end{cases}
\end{align}
 In other words, if all the matched filter outputs are smaller than $\eta_{\rm BT}^{(I)}$, $Y{=}\emptyset$ indicates that no beam pair is deemed sufficient for data transmission, either due to blockage ($B_{k}^{(I)}{=}0$), or 
 the UE receiving on the sidelobes of the BPIs $j\in{\mathcal S}_{\rm BT}$.
Otherwise, $Y{=}j^*$ indicates the index of the strongest BPI detected.

We now perform a probabilistic analysis of feedback. To this end,
let $S_I{=}s_I^*(X_k)$ and $B_I{=}B_k^{(I)}$ be the
SBPI and blockage state under BS $I$ at the beginning of the BT phase.
We assume that these state variables do not change during the transmission of the beacon sequences, i.e.,
$s_I^*(X_{k+i}){=}S_I$ and $B_{k+i}^{(I)}{=}B_I,{\forall} i{\in}\{0,{\ldots},T_{\rm BT}{-}2\}$. This is a reasonable assumption, since the duration of the BT phase ($\times$0.1ms) is typically much shorter than the time required by the UE to change beam ($\times$100ms) or the time-scales of blockage ($\times$100ms).
 With this assumption,  given the state $(S_I,B_I)$ of BS $I$ during BT, the signal sequence
$\{\Gamma_{j},{\forall} j{\in}{\mathcal S}_{\rm BT}\}$ is independent across $j$, 
due to the i.i.d. nature of $h_{k+i}^{(I)}$, $\Omega_{k+i}^{(I)}$ and $\mathbf w_{k+i}$. In addition,
in case of active  SBPI  ($S_I{=}j$ and $B_I{=}1$), by using \eqref{sigmodel} and \eqref{eq:PSNR}, $\Gamma_{j}$ has exponential distribution with mean
$1{+}{\rm SNR}_{\rm BT} L$,
 $\Gamma_{j}{\sim}\mathcal E(1{+}{\rm SNR}_{\rm BT}L )$; otherwise (inactive  SBPI, $S_I{\neq}j$ or $B_I{=}0$)
$\Gamma_{j}{\sim}\mathcal E(1{+}\rho_I{\rm SNR}_{\rm BT}L)$.
It follows that 
\begin{align*}
\begin{cases}
\Sigma_{I,1}\triangleq\mathbb P({\Gamma_{j}}{\leq}\eta_{\rm BT}^{(I)}|
S_I=j,B_I=1)
=1-e^{\frac{-\eta_{\rm BT}^{(I)}}{1+{\rm SNR}_{\rm BT} L}},\\
\Sigma_{I,0}\triangleq\mathbb P({\Gamma_{j}}{\leq}\eta_{\rm BT}^{(I)}|
S_I\neq j\text{ or }B_I=0){=}
1{-}e^{\frac{-\eta_{\rm BT}^{(I)}}{1+\rho_I{\rm SNR}_{\rm BT} L}}.
\end{cases}
\end{align*}
 Now, let us consider separately the two events
\mbox{$\{S_I\notin{\mathcal S}_{\rm BT}\}\cup \{ B_I=0\}$} (``inactive SBPI in ${\mathcal S}_{\rm BT}$'') and \mbox{$\{S_I\in{\mathcal S}_{\rm BT}\}\cap\{ B_I=1\}$} (``active SBPI $S_I \in {\mathcal S}_{\rm BT}$'').
In case of inactive  SBPI  in ${\mathcal S}_{\rm BT}$,
 the probability of generating the feedback signal $Y=\emptyset$ (i.e., of correctly detecting inactive  SBPI  within the $\mathcal S_{\rm BT}$ scanned in the BT phase) is
\begin{align}
\label{eq:fbbt1}
&\mathbb P(Y=\emptyset|\text{inactive } {\rm SBPI} \text{ in }{\mathcal S}_{\rm BT})
\\ 
&=
\prod_{{j}\in{\mathcal S}_{\rm BT}}\mathbb P({\Gamma_{j}}\leq \eta_{\rm BT}^{(I)}|{S_I\neq j\text{ or }B_I=0})
=
\Sigma_{I,0}^{|{\mathcal S}_{\rm BT}|},
\nonumber
\end{align}
since $Y{=}\emptyset$ is equivalent to ${\Gamma_{j}}\leq \eta_{\rm BT}^{(I)},\forall j\in{\mathcal S}_{\rm BT}$,
and $\Gamma_{j}$ are independent across $j$, conditional on $(S_I,B_I)$.
Similarly, in case of active  SBPI  $S_I \in {\mathcal S}_{\rm BT}$, the probability of incorrectly detecting inactive  SBPI is
\begin{align*}
\label{misdetection}
&\mathbb P(Y=\emptyset|\text{active } {\rm SBPI}\ S_I \in {\mathcal S}_{\rm BT})
\\ &=
\mathbb P({\Gamma_{j}}{\leq}\eta_{\rm BT}^{(I)}|{S_I{=}j,B_I{=}1})\!
\!\!\!\prod_{{j}\in{\mathcal S}_{\rm BT}{\setminus\{S_I\}}} \!\!\!\!\mathbb P({\Gamma_{j}}{\leq}\eta_{\rm BT}^{(I)}|{S_I{\neq} j,B_I{=}1})
\\ &
=
\Sigma_{I,1}\Sigma_{I,0}^{|{\mathcal S}_{\rm BT}|-1},
\numberthis
\end{align*}
since $S_I$ is the SBPI, implying $\Gamma_{s_I}\sim\mathcal E(1+{\rm SNR}_{\rm BT}L )$.
\\
\indent In case of inactive  SBPI in ${\mathcal S}_{\rm BT}$, the probability of generating the feedback signal $j^* \in \mathcal S_{\rm BT}$ (i.e., of incorrectly detecting an active SBPI) is
\begin{align*}
\label{falsealarm}\numberthis
&\mathbb P(Y={j}^*|\text{inactive } {\rm SBPI} \text{ in }{\mathcal S}_{\rm BT})\\
& {=}\frac{1}{|{\mathcal S}_{\rm BT}|}
\Bigr[1{-}\mathbb P(Y{=}\emptyset|\text{inactive }{\rm SBPI} \text{ in }{\mathcal S}_{\rm BT})\Bigr]
{=}\frac{1{-}\Sigma_{0,I}^{|{\mathcal S}_{\rm BT}|}}{|{\mathcal S}_{\rm BT}|};
\end{align*}
in fact, $\Gamma_{j}$ are i.i.d. across beams, conditional on inactive  SBPI, so that incorrect detections are uniform across the feedback outcomes ${j}^*\in{\mathcal S}_{\rm BT}$.
\\
\indent Instead, in case of active  SBPI $S_I \in {\mathcal S}_{\rm BT}$, we need to further distinguish between the two cases $j^* = S_I$ (the  SBPI  is detected correctly) and $j^*\in \mathcal S_{\rm BT}{\setminus} \{S_I\}$ (incorrect detection).
The probability of correctly detecting the  SBPI  is found as
\begin{align*}
\label{eq:fbbt2}
\nonumber
&\mathbb P(Y=S_I|\text{active } {\rm SBPI}\ S_I\in{\mathcal S}_{\rm BT})
\\ 
&{=}
\mathbb P(\Gamma_{S_I}{>}\eta_{\rm BT}^{(I)},\Gamma_{S_I}{>}\Gamma_{j}, \forall j{\in}{\mathcal S}_{\rm BT}{\setminus\{S_I\}}|\text{active } {\rm SBPI}\ S_I{\in}{\mathcal S}_{\rm BT})
\\ 
&{=}
\int_{\eta_{\rm BT}^{(I)}}^\infty \Bigl[f(\Gamma_{S_I}=\tau|\text{active } {\rm SBPI}\ S_I\in{\mathcal S}_{\rm BT})\\
&\qquad \times
\prod_{j\in{\mathcal S}_{\rm BT}{\setminus\{S_I\}}}
\mathbb P(\Gamma_{j}<\tau|S_I\neq j,B_I=1)\Bigr]\mathrm d\tau
\\
&
{=}
\int_{\eta_{\rm BT}^{(I)}}^\infty\biggl[
\frac{1}{1+{\rm SNR}_{\rm BT} L}\exp\left\{-\frac{\tau}{1+{\rm SNR}_{\rm BT} L}\right\}
\\
&\qquad\times\left(1-\exp\left\{-\frac{\tau}{1+\rho_I{\rm SNR}_{\rm BT} L}\right\}\right)^{|{\mathcal S}_{\rm BT}|-1} \biggr] \mathrm d\tau
\end{align*}\begin{align*}
&
{=}
\!\!\sum_{n=0}^{|{\mathcal S}_{\rm BT}|{-}1}
\left(\!\!\begin{array}{c}|{\mathcal S}_{\rm BT}|{-}1 \\ n\end{array}\!\!\right)
\frac{(-1)^{n}(1{-}\Sigma_{I,1})
(1{-}\Sigma_{I,0})^n}
{1{+}\frac{1+{\rm SNR}_{\rm BT} L}{1{+}\rho_I{\rm SNR}_{\rm BT} L}n}
, \numberthis
\end{align*}
where in the first step we used the definition of $Y{=}S_I$, i.e., $\Gamma_{S_I}$ must be greater than the threshold $\eta_{\rm BT}^{(I)}$, and all other $\Gamma_{j}$ must be smaller than $\Gamma_{S_I}$; in the last step, we used Newton's binomial theorem to solve the integral. Finally, the probability of incorrectly detecting the  SBPI, $j^* \in \mathcal S_{\rm BT} {\setminus} \{S_I \}$ is
\begin{align*}
\label{eq:fbbt3}
&\mathbb P(Y=j^*|\text{active  } {\rm SBPI}\ S_I\in{\mathcal S}_{\rm BT})\\
&{=}
\frac{1}{|{\mathcal S}_{\rm BT}|{-}1}\biggl[
1{-}\!\!\!\!\sum_{y\in\{S_I,\emptyset\}}\mathbb P(Y{=}y|\text{active  } {\rm SBPI}\ S_I{\in}{\mathcal S}_{\rm BT})
\biggr]\!\!\!\numberthis
\end{align*}
since, similarly to \eqref{falsealarm}, erroneous detections are uniform across the remaining $|{\mathcal S}_{\rm BT}|-1$ beams.
\\
\indent Since $Y{=}\emptyset$ represents the fact that the inactive SBPI condition has been detected, we choose $\eta_{\rm BT}^{(I)}$ so that the misdetection and false alarm probabilities are both equal to $\delta_{\rm BT}$, yielding from \eqref{misdetection}-\eqref{falsealarm} (over all $j{\in}{\mathcal S}_{\rm BT}$),
\begin{align}
\delta_{\rm BT}^{(I)}
=
1{-}\Sigma_{I,0}^{|{\mathcal S}_{\rm BT}|}
=
\Sigma_{I,1}\Sigma_{I,0}^{|{\mathcal S}_{\rm BT}|-1}.
\label{eq:deltaBT}
\end{align}
For a given ${\rm SNR}_{\rm BT}$ and 
$|{\mathcal S}_{\rm BT}|$, the value of $\eta_{\rm BT}^{(I)}$ and the corresponding $\delta_{\rm BT}^{(I)}$  can be found numerically using the bisection method, since the left- and right- hand sides of \eqref{eq:deltaBT} are decreasing and increasing functions of $\eta_{\rm BT}^{(I)}$, respectively.
\\
\indent \textbf{DT phase:} At the start of the DT phase, the BS $I$ chooses a BPI $j{\in}\mathcal S_I$
used for data transmission, along with the duration $T_{\rm DT}$ of the DT frame, the target
average SNR at the receiver ${\rm SNR}_{\rm DT}$, and a target transmission  rate $\bar R_{\rm DT}$; the last slot is used for the feedback signal from the UE to the BS, as described below. We assume that a fixed fraction $\kappa{\in}(0,1)$ out of $L$ symbols in each slot is used for channel estimation. 
Consider slot $t\in\{k,\dots, k+T_{\rm DT}-2\}$ of data communication; then,
if  $s_I^*(X_{t}){\neq} j$ or $B_{t}^{(I)}{=}0$, i.e., the selected BPI $j$ is inactive,
then the communication is in outage; otherwise
($s_I^*(X_{t}){=}j$ and $B_{t}^{(I)}{=}1$, i.e., the selected BPI $j$ is an active SBPI)
 assuming that channel estimation errors are negligible compared to the noise level (achieved with a sufficiently long pilot sequence $\kappa L$),
from the signal model \eqref{sigmodel},
 we find that outage occurs if (note that $\mathbb E[|h_t^{(I)}|^2\ell(d_I(X_t))]=1$) 
\begin{align} 
W_{\rm tot} \log_2 (1+|h_t^{(I)}|^2\ell(d_I(X_t)){\rm SNR}_{\rm DT})<\bar R_{\rm DT},
\end{align}
yielding the outage probability
\begin{align} \label{eq:outprob}
\nonumber
\mathbb P_{\text{OUT}}(\bar R_{\rm DT},{\rm SNR}_{\rm DT})&{=}
\mathbb P\Bigr(|h_t^{(I)}|^2\ell(d_I(X_t)){<}\frac{2^{\bar R_{\rm DT}/W_{\rm tot}}{-}1}{{\rm SNR}_{\rm DT}}\Bigr)\\
&{=}
1{-}\exp\Bigr\{-\frac{2^{\bar R_{\rm DT}/W_{\rm tot}}-1}{{\rm SNR}_{\rm DT}}\Bigr\}.
\end{align}
 In this paper, we design $\bar R_{\rm DT}$ based on the notion of $\epsilon-$outage capacity, i.e., $\bar R_{\rm DT}$ is the largest rate such that $\mathbb P_{\text{OUT}}(\bar R_{\rm DT},{\rm SNR}_{\rm DT}) \le \epsilon$, for a target outage probability $\epsilon<1$. Imposing \eqref{eq:outprob} equal to $\epsilon$, this can be expressed as 
 \begin{align} \label{eq:R_t}
\!\!\bar R_{\rm DT}{=}C_{\epsilon}({\rm SNR}_{\rm DT}){=} W_{\rm tot} \log_2\left(1-{\rm SNR}_{\rm DT} \ln(1-\epsilon)\right).\!\!
\end{align}
With this choice, the transmission is successful with probability $1-\epsilon$, and the average rate (throughput) is 
\begin{align}
\label{thr}
\mathcal T(\epsilon,{\rm SNR}_{\rm DT})\triangleq(1-\kappa)(1-\epsilon)C_{\epsilon}({\rm SNR}_{\rm DT}),
\end{align}
where $(1-\kappa)$ accounts for the channel estimation overhead.  
In what follows, we select $\epsilon$ to maximize the throughput, yielding  the optimal $\epsilon^*({\rm SNR}_{\rm DT})$ at a given SNR ${\rm SNR}_{\rm DT}$ as the unique fixed point of $\mathrm d \mathcal T(\epsilon,{\rm SNR}_{\rm DT})/\mathrm d\epsilon=0$, or equivalently, $$\ln\Big(1-{\rm SNR}_{\rm DT}\ln(1-\epsilon)\Big)\Big(1-{\rm SNR}_{\rm DT}\ln(1-\epsilon)\Big)={\rm SNR}_{\rm DT}.$$
 We denote the resulting throughput maximized over $\epsilon$ as 
 $\mathcal T^*({\rm SNR}_{\rm DT})\triangleq \mathcal T(\epsilon^*({\rm SNR}_{\rm DT}),{\rm SNR}_{\rm DT})$.
\\
\indent 
We envision a mechanism in which the pilot signal transmitted in the last data transmission slot (at time $t=k+T_{\rm DT}-2$) is used to generate the binary feedback signal
\begin{align}
\label{fbdt}
 Y = \begin{cases}
{j}, &{\Gamma_{j}}>\eta_{DT}^{(I)},\\
{\emptyset}, &{\Gamma_{j}}\leq\eta_{DT}^{(I)},
  \end{cases}
  \end{align}
 transmitted by the UE to the BS in the last slot of the DT phase (at time $t=k+T_{\rm DT}-1$). 
 As in \eqref{zsi} for the BT feedback,  $Y{=}j$ denotes
active  SBPI  detected, whereas $Y=\emptyset$ denotes inactive SBPI detection, due to either loss of alignment or blockage.
Similarly to \eqref{zsi}, 
  $$\Gamma_{j}\triangleq\frac{|\mathbf{x}_{k+T_{\rm DT} -2}^{(p)H} \mathbf{y}_{k+T_{\rm DT} -2}^{(p)}|^2}{(1+F)N_0 W_{\rm tot} \Vert\mathbf{x}_{k+T_{\rm DT} -2}^{(p)}\Vert_2^2}$$
is based on the pilot signal $\mathbf{x}_{k+T_{\rm DT} -2}^{(p)}$ (of duration $\kappa L$)
  and on the corresponding signal $\mathbf{y}_{k+T_{\rm DT} -2}^{(p)}$ received on the second last slot of the DT phase.
  The distribution of the feedback conditional on
  $s_I^*(X_{t}){=}S_I$ and $B_{t}^{(I)}{=}B_I$ at the 2nd last slot ($t{=}k{+}T_{\rm DT}{-}2$)
 can be computed as a special case of \eqref{misdetection} and \eqref{falsealarm} with $|{\mathcal S}_{\rm BT}|=1$ (since in the DT phase only one beam $j$ is used for data transmission) and $\kappa L$ in place of $L$ (since only  $\kappa L$ symbols are used as pilot signal), yielding the probability of incorrectly detecting an active  SBPI   as
    \begin{align}
    \label{eq:dtfa}
&\!\!\mathbb P(Y{=}j|
{S_I\neq j\text{ or }B_I=0}){=}\!\exp\!\left\{{-}\frac{\eta_{DT}}{1{+}\rho_I\kappa{\rm SNR}_{\rm DT} L}\right\}\!\!,\!\!
\end{align}
and that of incorrectly detecting $j$ to be an 
inactive  SBPI as
\begin{align}
 \label{eq:dtmd}
&\!\!\!\mathbb P(Y{=}\emptyset|{S_I=j,B_I=1}
){=}1{-}\exp\left\{-\frac{\eta_{DT}}{1{+}\kappa{\rm SNR}_{\rm DT} L}\right\}.\!\!
\end{align}
As in the BT phase, we choose $\eta_{DT}^{(I)}$ so that the probabilities of misdetection and false alarm are both  equal to $\delta_{\rm DT}^{(I)}$, yielding
\begin{align}
\!\!\!\delta_{\rm DT}^{(I)}{=}\!\exp\!\left\{\!\frac{-\eta_{DT}}{1{+}\rho_I\kappa{\rm SNR}_{\rm DT} L}\!\right\}\!{=}
1{-}\exp\!\left\{\!\frac{-\eta_{DT}}{1{+}\kappa{\rm SNR}_{\rm DT} L}\!\right\}\!{.}\!\!\!
\label{eq:deltat}
\end{align}

\section{POMDP Formulation}
\label{sec:POMDP}
We now formulate the problem of optimizing the BT, DT and HO strategy as a 
 constrained POMDP.
 In the following, we define the elements of this POMDP.

\noindent \underline{\textbf{States:}} the state at time $k$ is denoted by $Z_k$. We introduce the state $\bar z$ to characterize the episode termination, so that $Z_k{=}\bar z$ if the UE exited the coverage area of the two BSs, i.e., $X_k{\notin}\mathcal X$. Otherwise ($Z_k{\neq} \bar z$), we define the state as $Z_k {\triangleq }(U_k,I_k)$, where $I_k{\in}\mathcal I$ is the index of the serving BS, $U_k {\triangleq} (S_k,B_k)$ is the joint SBPI-blockage state, taking values from the set $\mathcal U{=}\mathcal S {\times}\{0,1\}^2$, ${S_k}{=}(S_k^{(0)},S_k^{(1)}){\in} \mathcal S$ with $S_k^{(i)}{\triangleq} s^*_i(X_k)$ is the SBPI at the current UE position $X_k$, $B_k{ = }(B_k^{(0)},B_k^{(1)})$ is the blockage state of the two BSs.
The overall state space, including the absorbing $\bar z$, is then $\mathcal{Z}{=} (\mathcal U \times \mathcal I)\cup\{\bar z\}$. \label{partial_obs} Note that the position of the UE and the blockage state cannot be directly observed, thereby making the state $U_k$ unobservable. We model such state uncertainty via a belief $\beta_k$, representing the probability distribution of $U_k$, given the information collected (actions selected and feedback) up to time $k$.

\noindent \underline{\textbf{Actions:}} the serving BS can perform three actions: beam-training (BT), data transmission (DT), or handover (HO). However, differently from standard POMDPs in which each action takes one slot, in this paper we generalize the model to actions taking multiple slots, as explained next.

If action HO is chosen, the data plane is transferred to the other BS, which becomes the serving one for the successive  \mbox{time-slots}, until HO is chosen again or the episode terminates. HO requires $T_{\rm HO}$  \mbox{time-slots} to complete, due to the delay to coordinate the transfer of the data traffic between the two BSs.

If actions BT is chosen, the serving BS $I$ chooses the BPI set ${\mathcal S}_{\rm BT} {\subseteq} \mathcal S_I$
 to scan and the target SNR {$\mathrm{SNR}_{\mathrm{BT}}$}. The transmission power is then found via \eqref{eq:PSNR}, and the feedback error probability $\delta_{\rm BT}^{(I)}$ is found by solving \eqref{eq:deltaBT}.
The action duration is $T_{\rm BT}=|{\mathcal S}_{\rm BT} |+1$: $|{\mathcal S}_{\rm BT} |$ slots for 
scanning the BPI set ${\mathcal S}_{\rm BT} $, and one slot for the  feedback back to the serving BS.

If action DT is chosen, then the serving BS $I$ selects the BPI  $j {\in} \mathcal S_I$ to perform data communication with the UE, along with the duration $T_{\rm DT}{\geq} 2$ of the data communication session, and the target SNR $\mathrm{SNR}_{\mathrm{DT}}$.
The transmission power is then determined via \eqref{eq:PSNR}, and the transmission rate is given by \eqref{eq:R_t} to achieve $\epsilon$-outage capacity, so that the resulting throughput (in case of LOS and correct alignment) is $\mathcal T^*({\rm SNR}_{\rm DT})$.
The duration of the data communication session $T_{\rm DT}$ includes the second last slot for the feedback signal, which is transmitted from the UE to the BS in the last slot.
The feedback error probability $\delta_{\mathrm{DT}}^{(I)}$ is the unique fixed point of \eqref{eq:deltat}.

We represent compactly these actions as
$(c, \Pi_c){\in\mathcal A_I}$, with action space $\mathcal A_I$, where $c{\in}\{ {\mathrm{BT}}, {\mathrm{DT}}, {\mathrm{HO}} \}$ refers to the action class and $\Pi_c{=}({\mathcal S}_c, \mathrm{SNR}_c, T_c)$ specifies the corresponding  parameters: ${\mathcal S}_c\subseteq\mathcal S_I$ is a subset of BPIs of serving BS $I$, used during the action, $\mathrm{SNR}_c$ is the target SNR, so that the corresponding transmission power is given by \eqref{eq:PSNR}, and $T_c$ is the action duration.
For HO, we set $\mathrm{SNR}_{\rm HO} {=} 0$ and $\mathcal S_{\rm HO}{=}\emptyset$.

	\noindent \underline{\textbf{Observations:}}
		after selecting action $A_k{\in}\mathcal A_I$ of duration $T$ in slot $k$ and executing it in state $u_k{\in}\mathcal U$, the BS observes $Y_{k+T}$ taking value from the observation space $\bar {\mathcal Y} \triangleq  \mathcal Y {\cup}\{\bar{z}\} $, where $\mathcal Y\triangleq{\mathcal S_1}{\cup}{\mathcal S_2}{\cup}\{\emptyset\}{\cup}\{\bar z\}$.
		  $Y_{k+T}{=}\bar{z}$ denotes that $Z_{k}=\bar z$, so that the UE exited the coverage area of the two BSs and the episode terminates; otherwise, $Y_{k+T}$ denotes the feedback signal after the action is completed, as described in \eqref{btfb} and \eqref{fbdt} for the BT and DT actions (${Y}_{k}{=}\emptyset$ under the HO action).
\\
	\noindent \underline{\textbf{Transition and Observation probabilities:}}
	\label{sec:state_obs_prob}
	 Let $\mathbb P({Z_{k+T}=z'},Y_{k+T}=y|{Z_k=z},A_k=a)$ be the probability of moving from a \mbox{non-absorbing} state $z=(u,I)\in\mathcal Z\setminus\{\bar z\}$ to state $z'\in\mathcal Z$ and observing $y\in\bar{\mathcal Y}$ 
under action $a\in \mathcal A_I$ of duration $T$.
If the episode does not terminate ($Z_{k+T}\neq \bar z$ and $y\neq \bar z$),
let $Z_{k+T}=(u',I')$ be the next state.
Note that the new serving BS $I'$ is a function $\mathbb I(a,I)$ of the chosen action:
if $a$ is the HO action  then $I'=\mathbb I(a,I)=1-I$, otherwise $I'{=\mathbb I(a,I)}=I$. 
Using the law of conditional probability,
the transition probability is then expressed as
\begin{align}
\label{eq:UIsep}
&\mathbb P(Z_{k+T}{=}(u',I'),Y_{k+T}{=}y|Z_k{=}(u,I),A_k{=}a)
\\
&=
\mathbb P(U_{k+T}{=}u',Y_{k+T}{=}y|
U_k{=}u,I_k{=}I,A_k{=}a)
\chi(I' {=} \mathbb I(a,I)),
\nonumber
\end{align}
since $(U_{k+T},Y_{k+T})$ is conditionally independent of $I_{k+T}$ given $(U_k,I_k,A_k)$.
 To characterize the first term in \eqref{eq:UIsep},
under the HO action $a{=}({\mathrm{HO}},\emptyset, 0, T_{\rm HO})$, of duration $T{=}T_{\rm HO}$, the observation signal is deterministically $Y_{k+T}{=}\emptyset$,
yielding
\begin{align}\nonumber
&\mathbb P(U_{k+T}{=}(s'{,}b'){,}Y_{k+T}{=}\emptyset|U_k{=}(s{,}b){,}I_k{=}I,A_k{=}a)\\
&\qquad= \mathbf P_{s'b'|sb}(T),
\label{eq:uoprob_ho}
\end{align}
where $\mathbf P_{s'b'|sb}(T)$ is the $T$ steps transition probability from
$U_k{=}(s,b)$ to $U_{k+T}{=}(s',b')$, found recursively as 
$\mathbf P_{s'b'|sb}(T){=}\sum_{s'',b''}
\mathbf P_{s'b'|s''b''}(T{-}1)\mathbf P_{s''b''|sb}$ with $
\mathbf P_{s'b'|sb}(1){=}\mathbf P_{s'b'|sb}$. In other words,
the UE moves from $s$ to $s'$ and the BSs' blockage states  move from $b$ to $b'$, in $T$ slots.
\\
\indent
Under the BT action $a{=}({\mathrm{BT}},{\mathcal S}_{\rm BT},{\rm SNR}, T)$, of duration
$T{=}|{\mathcal S}_{\rm BT}|+1$, the observation signal is $Y_{k+T}=y\in{\mathcal S_{\rm BT}}\cup\{\emptyset\}$ (see the BT signaling mechanism in \secref{sec:System_Model}). Therefore, 
\begin{align*}
&\mathbb P\Big(U_{k+T}=(s',b'),Y_{k+T}=y|U_k=(s,b),I_k=I,A_k=a\Big)
\\ 
&=
\mathbb P(Y_{k+T}{=}y|{\mathcal S}_{\rm BT},S_k^{(I)}{=}s_I,B_{k}^{(I)}{=}b_I,I_k{=}I)\mathbf P_{s'b'|sb}(T){,}
\end{align*}
where $\mathbb P(Y{=}y|{\mathcal S},S_k^{(I)}{=}s_I,B_{k}^{(I)}{=}b_I,I_k{=}I)$ has been defined in \eqref{eq:fbbt1}-\eqref{eq:fbbt3}
for the cases of active  SBPI  $\{s_I{\in}{\mathcal S}\}\cap\{b_I = 1\}$ and inactive  SBPI  $\{s_I{\notin}{\mathcal S}\}\cup\{b_I = 0\}$.
\\\indent
Finally, under the DT action $a{=}({\mathrm{DT}},\{j\},{\rm SNR}, T)$, the observation signal is $Y_{k+T}{=}y\in\{j,\emptyset\}$ (see the DT signaling in \secref{sec:System_Model}). However, in this case the feedback signal is generated based on the second last slot, i.e., it depends on the state $U_{k+T-2}$ at time $k{+}T{-}2$.
By  marginalizing with respect to $S_{k+T-2}{=}s''$ and $B_{k+T-2}{=}b''$, we then obtain \eqref{eq:uoprob_dt} given at the top of page \pageref{eq:uoprob_dt}.
\begin{figure*}[!t]
\centering
\begin{align*}
\label{eq:uoprob_dt}
\numberthis
&\mathbb P\Bigr(U_{k+T}=(s',b'),Y_{k+T}=y|U_k=(s,b),I_k =I, A_k=a\Bigr)
\\
&=
\sum_{s''\in\mathcal S,b''\in\{0,1\}^2}\!\!\!\!\!\!\!\!\mathbb P\Bigr(U_{k{+}T}{=}(s',b'){,}Y_{k+T}{=}y,S_{k{+}T{-}2}{=}s''{,}B_{k{+}T{-}2}{=}b''
|U_k{=}(s,b),I_k=I,A_k=a\Bigr)\\
&=
\sum_{s''\in\mathcal S,b''\in\{0,1\}^2} \biggl[
\mathbf P_{s''b''|sb}(T-2)
\mathbb P\Bigr(Y_{k+T}{=}y|\{j\}, S_{k+T-2}^{(I)}{=}s_I'',B_{k+T-2}^{(I)}{=}b_I'',I_{k+T-2}{=}I\Bigr)
\mathbf P_{s'b'|s''b''}(2)
\biggr] \nonumber\\
\hline
\end{align*}
\vspace{-15mm}
\end{figure*}
To explain it, note that:
the system moves from $(S_{k},B_{k}){=}(s,b)$ to 
$(S_{k+T-2},B_{k+T-2}){=}(s'',b'')$ in $T{-}2$ steps;
then, the feedback signal $Y_{k+T}$ is generated
with distribution $\mathbb P(Y_{k+T}{=}y|\{j\}, S_{k+T-2}^{(I)}{=}s_I'',B_{k+T-2}^{(I)}{=}b_I'',I_{k+T-2}{=}I)$, given in \eqref{eq:dtfa}, \eqref{eq:dtmd}
for the cases of active or inactive  SBPI  in $\{j\}$;
finally, in the remaining $2$ steps ,the system moves from 
$(S_{k+T-2},B_{k+T-2}){=}(s'',b'')$ to
$(S_{k+T},B_{k+T}){=}(s',b')$.
\\\indent
The probability of terminating the episode ($z'=\bar z$ and $y=\bar z$) is equivalent
to the probability of exiting the coverage area of the two BSs within $T$ steps,
\begin{align*}
&\mathbb P(Z_{k+T}{=}\bar{z},Y_{k+T}{=}\bar{z}|Z_k=(u,I),A_k{=}a)\\
&{=}
1{-}\!\!\!\!\!\sum_{u'\in\mathcal U,y\in \mathcal Y} \!\!\!\!\!\mathbb P(U_{k+T}{=}u',Y_{k+T}=y|
U_k{=}u,I_k{=}I{,}A_k{=}a)
\end{align*}
since it
 is the complement event of \mbox{$\cup_{z\in \mathcal Z \setminus\{ \bar z\}}\cup_{y\in \mathcal Y}\{ Z_k = z, Y_{k+T}=y\}$}.
\\
\noindent \underline{\textbf{Costs and Rewards:}}
for every state $z = (u,I)\in\mathcal Z\setminus\{\bar z\}$ and action $a$, we let
$r(u,I,a)$ and $e(u,I,a)$ be the expected number of bits transmitted from the serving BS to the UE and the expected energy  cost, respectively.
Under the HO and BT actions, we have that $r(u,I,a)=0$ (since no bits are transmitted during these actions).
On the other hand, under the DT action
$a=(\mathrm{DT},\{j\}, \mathrm{SNR}, T_{\rm DT})$ taken  in slot $k$, the expected throughput in the $t$th communication slot, $t{\in}\{0,\dots, T_{\rm DT}-2\}$, is $\mathcal T^*({\rm SNR})$ as in \eqref{thr}, maximized over $\epsilon$, if the current state is such that $S_{k+t}^{(I)}{=}j$ and 
$B_{k+t}^{(I)}=1$ (i.e., $j$ is an active  SBPI); otherwise, outage occurs and the expected throughput is zero. Therefore, we find that
\begin{align}
\nonumber
&r((s,b),I,(\mathrm{DT},\{j\}, \mathrm{SNR}, T_{\rm DT}))\\ \nonumber
&
{=}\mathcal T^*({\rm SNR})
\sum_{t=0}^{T_{\rm DT}-2}\mathbb P(S_{k+t}^{(I)} {=} j,B_{k+t}^{(I)}{=}1|S_k{=}s,B_{k}{=}b)
\\  &
{=}
\mathcal T^*({\rm SNR})
\sum_{t=0}^{T_{\rm DT}-2} \sum_{(s',b')\in \mathcal U}
{\mathbf P_{s'b'|sb}(t)}\chi(s_I' = j,b_I' = 1).\!\!\!
\label{eq:reward}
\end{align}
\indent The energy cost of a HO action is $e(u,I,a){=}0$; that of DT or BT action 
$a{=}(c,{\mathcal S}, {\rm SNR},T )$ is found from \eqref{eq:PSNR} as
(note that $T{=}|{\mathcal S}|{+}1$ for a BT action and $|{\mathcal S}|{=}1$ for a DT action)
\begin{align}
e(u,I,a)
=
\frac{(T-1)\Delta_t}{|{\mathcal S}|}\sum_{j\in {\mathcal S}} \frac{\sigma_w^2}{\Upsilon_{j,I}+\sigma_{{\rm DIF},I}^2} {\rm SNR}.
\label{eq:cost}
\end{align}
Note that the last slot of the DT or BT phases is reserved to the feedback transmission, with no energy cost for the BS.

\noindent \underline{\textbf{Policy and Belief updates:}}
Since the agent cannot directly observe the pairs of BPI $S$ and blockage $B$, we define the POMDP state as $(\beta,I)$, where  
 $\beta$ denotes the belief, i.e., the probability distribution over $U{=}(S,B)$, given the information collected so far and $I$ is the index of the serving BS. The belief $\beta$ takes values from belief  space $\mathcal B {\triangleq} \{ \beta{\in} \mathbb R^{|\mathcal U|} : \beta(u)\geq 0\ \forall u{\in}\mathcal U, \sum_{u\in\mathcal U} \beta(u) =1\}$.
Given $(\beta,I)$, the serving BS selects an action $a$ according to a policy $a=\pi(\beta,I)$, that is part of our design in \secref{sec:Problem_Optimization}; then, after executing the action $a$ and receiving the feedback signal $y \in \mathcal Y$, the BS $I$ updates the belief according to Bayes' rule as
\begin{align}
\nonumber
&\beta'(u') {=}\mathbb P(u'\!\mid\!y, a, \beta,I)
\\&
 {=} \frac{\sum_{u \in \mathcal{U}} 
\beta(u)\mathbb P(u',y|u,I,a)}{
\sum_{u\in \mathcal{U}}\beta(u)
{\sum_{u''\in \mathcal{U}}\mathbb P(u'',y|u,I,a)}
},
\label{eq:belief_update}
\end{align}
with $\mathbb P(u',y |u,I,a)$
given by \eqref{eq:uoprob_ho}-\eqref{eq:uoprob_dt}, and the serving BS becomes $I'=\mathbb I(a,I)$.
We denote the function that maps the belief $\beta$, action $a$ and observation $y$ under the serving BS $I$ as 
$\beta'=\mathbb B_I(y{,}a{,}\beta)$. Note that $Y{=}\bar z$ indicates episode termination.
\section{Optimization Problem}
\label{sec:Problem_Optimization}
Our goal is to determine a policy $\pi$ (a map from beliefs to actions) maximizing the 
expected throughput, under an average power constraint $\bar P_{\rm avg}$, starting from an initial belief
$\beta_0{=}\beta_0^*$ and serving BS $I_0{=}I_0^*$.  
From Little's Theorem \cite{LittlesTheorem}, the average rate and power consumption
can be expressed as
\begin{align}
\label{metrics}
\bar T^{\pi}\triangleq \frac{\bar R_{\rm{tot}}^{\pi}}{\bar D_{\rm{tot}}},\ 
\bar P^{\pi}\triangleq \frac{\bar E_{\rm{tot}}^{\pi}}{\bar D_{\rm{tot}}},
\end{align}
where $\bar R_{\rm{tot}}^{\pi}$, $\bar E_{\rm{tot}}^{\pi}$  are the total 
expected number of bits transmitted and energy cost during an episode; $\bar D_{\rm{tot}}$
is the expected episode duration,
which only depends on the mobility process but is independent of the policy $\pi$.
Therefore, we aim to solve
\label{page:optprob}
\begin{align*}
&\textbf{P1:}\\
 &\!\!\max_\pi \, 
\bar R_{\rm{tot}}^{\pi}{\triangleq}
\mathbb{E}_{\pi}\! \bigg [ \!\sum_{n=0}^{\infty} r(u_{t_n},i_{t_n},a_{t_n}) \chi(Z_{t_n}{\neq}\bar z) \Big| \beta_0{=}\beta_0^*,I_0{=}I_0^* \bigg ], 
\\&
  \text{s.t. }\\
  &\bar E_{\rm{tot}}^{\pi}{\triangleq}\mathbb{E}_{\pi} \!\bigg [\! \sum_{n=0}^{\infty} \!e({u_{t_n}{,}i_{t_n}{,}a_{t_n}})  \chi(Z_{t_n}{\neq} \bar z) \Big| \beta_0{=}\beta_0^*{,}I_0{=}I_0^* \!\bigg ] {\le} E_{\max},
\end{align*}
where $E_{\max} {\triangleq} \bar D_{\rm{tot}}\bar P_{\rm avg}$; $t_n$ is the time index of the $n$-th decision round, recursively computed as $t_{n+1} {=} t_n {+} T_{n}$, where $T_{n}$ is the duration (number of slots) of the action taken in the $n$-th decision round and $t_0 {=} 0$. 
 We opt for a Lagrangian relaxation
to handle the cost constraint, and define
$\mathcal{L}_\lambda(u,i,a)=r(u,i,a){-}\lambda e(u,i,a)$ for $\lambda{\geq} 0$.
For a generic policy $\pi$, we define its value function as\footnote{{Note that the convergence of this series  is guaranteed
 by the presence of the absorbing state $\bar{z}$.}}
\begin{align*}
V_\lambda^\pi(\beta,I) {=} \mathbb{E}_{\pi} \bigg [ \sum_{n=0}^\infty \mathcal{L}_\lambda(u_{t_n},i_{t_n},a_{t_n}) \chi(Z_{t_n}{\neq} \bar z) \mid \beta_0{=}\beta,I_0{=}I \bigg ].
\end{align*}
The goal is to determine the optimal policy $\pi^*$ which maximizes the value function, i.e.,
\begin{align}
V_\lambda^*(\beta,I)\triangleq\max_\pi \, {V_\lambda^\pi(\beta,I)}.
\end{align}
The optimal dual variable is then found via the dual problem \begin{align}\label{dual}
\lambda^*=\arg\min_{\lambda\geq 0}V_\lambda^*(\beta_0^*,I_0^*)+\lambda E_{\max}.\end{align}
It is well known that the optimal value function for a given $\lambda$ uniquely satisfies Bellman's optimality equation~\cite{DBLP:journals/corr/abs-1109-2145}
$V_\lambda^*=H_\lambda[V_\lambda^*]$, where we have defined the operator $\hat V=H_\lambda[V]$ as
\begin{align*}
&\hat V(\beta,I) =\max_{a \in \mathcal{A}}\sum_{u \in \mathcal{{U}}} \beta(u)\bigg [
\mathcal{L}_\lambda(u,I, a) \\
&{+}\!\!\!\!\!\!\!\sum_{(u',y)\in \mathcal U \times \mathcal Y}\!\!\!\!\!\!\!\!\!
\mathbb P(u'{,}y|u{,}I,a) V\Big({\mathbb B_{{I}}(y{,}a{,}\beta){,}\mathbb I(a,I)}\Big)\bigg],
\ \forall (\beta,I){\in} \mathcal B \times \mathcal I.
\end{align*}
The optimal value function $V_\lambda^*$ can be arbitrarily well approximated via the value iteration algorithm $V_{n+1}{=}H_\lambda[V_n]$, where $V_0(\beta,I){=}0,{\forall}(\beta,I) \in \mathcal B \times \mathcal I$. 
Moreover, $V_n$ is
 a piece-wise linear and convex function~\cite{DBLP:journals/corr/abs-1109-2145}, so that, at any stage of value iteration, it can be expressed by a finite set of hyperplanes $\mathcal Q_n^{(I)}\equiv\{(\alpha_{n,I,\ell}^{(r)},\alpha_{n,I,\ell}^{(e)})\}_{\ell=1}^{N_n^{(I)}}$ of cardinality $N_n^{(I)}$, 
\begin{align}
\label{vn}
V_n(\beta,I)=\max_{\alpha_I\in\mathcal Q_n^{(I)}} 
\inprod{\beta}{ \alpha_I^{(r)}-\lambda\alpha_I^{(e)}},
\end{align}
 where $\inprod{\beta}{\alpha}=\sum_u \beta(u)\alpha(u)$ denotes inner product.
  Each hyperplane $(\alpha_I^{(r)},\alpha_I^{(e)})\in\mathcal Q_n^{(I)}$ is associated with an action $a_{\alpha_I}\in \mathcal{A}_I$,
 so that the maximizing hyperplane $\alpha_I^*$ in \eqref{vn} defines the policy $\pi_n(\beta,I)=a_{\alpha_I^*}$.
 Note that a distinguishing feature of our approach
 compared to \cite{DBLP:journals/corr/abs-1109-2145} is that we define distinct hyperplanes
 $\alpha_I^{(r)}$
  for the reward and $\alpha_I^{(c)}$ for the cost; as we will see later, this approach will be key to solving the dual optimization problem to optimize the power constraint, since it
  allows to more efficiently track changes in the dual variable $\lambda$, as part of the dual problem \eqref{dual},
 and to approximate the expected total reward and cost as
\begin{align}
\label{costrew} \nonumber
\bar R_n(\beta,I)= \inprod{\beta }{\alpha_I^{(r)*}},\ \bar E_n(\beta,I)=\inprod{\beta}{\alpha_I^{(e)*}},\\
\text{ where }
(\alpha_I^{(r)*},\alpha_I^{(e)*})=\arg \!\!\! \max_{\alpha_I\in\mathcal Q_n^{(I)}} \inprod{\beta}{\alpha_I^{(r)}-\lambda\alpha_I^{(e)}}.
\end{align}

It can be shown (see for instance \cite{Pineau2006PointbasedAF}) that the set of hyperplanes is updated recursively as
\begin{align*}
\label{setQ} 
&\mathcal Q_{n+1}^{(I)}
\equiv
\Bigl\{
(r(\cdot,I,a),e(\cdot,I,a))\\
&\qquad +
  \sum_{u'\in \mathcal U, y\in \mathcal Y}\mathbb P(u',y|\cdot,I,a)
  \left(\alpha_{I',y}^{(r)}(u'),\alpha_{I',y}^{(e)}(u')\right){:}\\
&a{\in}\mathcal A_I,
I' = \mathbb I(a,I),[(\alpha_{I',y}^{(r)},\alpha_{I',y}^{(e)})]_{\forall y\in\mathcal Y}{\in}\mathcal (Q_n^{(I')})^{|\mathcal Y|}
 \Bigr\},\numberthis
\end{align*}
so that the cardinality grows as $N_{n+1}^{(I)}=|\mathcal Q_{n+1}^{(I)}|
=\mathcal O(|\mathcal A|^{|\mathcal Y|^n})$ -- doubly exponentially with the number of iterations.

For this reason, computing optimal planning solutions for POMDPs is an intractable problem for any reasonably sized task.
 This calls for approximate solution techniques, e.g., PERSEUS~\cite{DBLP:journals/corr/abs-1109-2145}, which we introduce next.

PERSEUS \cite{DBLP:journals/corr/abs-1109-2145} is an approximate PBVI algorithm for unconstrained POMDPs. 
 Its key idea is to define an approximate backup operator $\tilde{H}_\lambda[\cdot]$ (in place of $H_{\lambda}[\cdot]$), restricted to a discrete subset of POMDP states in $ \tilde{\mathcal B}_0\cup\tilde{\mathcal B}_1$, where $\tilde{\mathcal B}_I$ is discrete set of POMDP states with the serving BS $I$, chosen as representative of the entire belief space $\mathcal B$;
 in other words, for a given value function $\tilde V_n$ at stage $n$, PERSEUS builds a value function $\tilde V_{n+1} {=} \tilde{H}[\tilde V_n]$ that improves the value of all POMDP states $(\beta,I)$ with $\beta\in \tilde{ \mathcal B}_I$,
without regard for the POMDP states outside of this discrete set, $\beta {\notin}\tilde{\mathcal B}_I$.
For each $ I \in \mathcal I$, the goal of the algorithm is to provide a $|\tilde{\mathcal{B}}_I|$-dimensional set of hyperplanes ${\alpha_{I}=(\alpha_{I}^{(r)},\alpha_{I}^{(e)})\in}\mathcal Q_{I}$ and associated actions $a_{\alpha_I}$.
Given such set,
the value function at any other POMDP state, $(\beta,I)$ is then approximated
via \eqref{vn}
as $\tilde V(\beta,I){=}\inprod{\beta}{\alpha_I^{(r)*}{-}\lambda\alpha_I^{(e)*}}$,  where $\alpha_I^*{=}(\alpha_I^{(r)*},\alpha^{(e)*}){=}\arg\max_{(\alpha_I^{(r)},\alpha^{(e)})\in\mathcal Q^{(I)}} \inprod{\beta}{\alpha_I^{(r)}{-}\lambda\alpha_I^{(e)}}$,
which defines an approximately optimal policy $\pi(\beta,I){=}a_{\alpha_I^*}$. 
\par Key to the performance of PBVI is the design of $\tilde{\mathcal{B}}_I$, which should be representative of the belief points encountered in the system dynamics.  In the PBVI literature~\cite{Pineau2006PointbasedAF}, most of the strategies to design $\tilde{\mathcal{B}}_I$ focus on selecting reachable belief points, rather than covering uniformly
 the entire belief simplex.
 We choose the beliefs in the following two steps. For each $I\in \mathcal I$, an initial belief set $\mathcal B_I^{(0)} $ is selected deterministically to cover uniformly the belief space.
  followed by expansion of $\{\mathcal B_I^{(0)},I\in\mathcal I\} $ using the \emph{Stochastic simulation and exploratory action} (SSEA) algorithm \cite{Pineau2006PointbasedAF} to yield the expanded belief points set $\{\tilde{\mathcal B}_I,I\in\mathcal I\}$.
After initializing $\mathcal{B}_I^{(0)}$, given $\mathcal{B}_I^{(n)}$ at iteration $n$,
 for each $\beta\in \mathcal{B}_I^{(n)}$, SSEA performs a one step forward simulation with each action in the action set, thus producing new POMDP states $\{(\beta_{a},I_a),\forall a\in\mathcal A_I\}$. At this point, it computes the L1 distance between each new  $\beta_a$ and its closest neighbor in $\mathcal{B}_{I_a}^{(n)}$, and adds the point $\beta_{a^*}$ to $\mathcal{B}_{I_{a^*}}^{(n)}$ if $\min_{\beta \in \mathcal B_{I_{a^*}}^{(n)}}\|\beta_{a^*} -\beta\|_1 \geq \min_{\beta \in \mathcal B_{I_{a}}^{(n)}}\|\beta_{a} -\beta\|_1, \forall a \in \mathcal A_I$, so as to more widely cover 
 the belief space. This expansion is performed multiple times to obtain $\{\tilde {\mathcal B}_I,I \in \mathcal I\}$.

The approximate backup operation of PERSEUS is given by Algorithm \ref{alg:alg_1},
which takes as input the index of the serving BS $I$, the set of belief points  $\tilde{\mathcal B}_I$ associated with BS $I$, the sets of hyperplanes $\{\mathcal Q_n^{(i)} ,i\in \mathcal I\}$ and the corresponding actions, and outputs a new set $\mathcal Q_{n+1}^{(I)}$ along with their corresponding actions.
To do so:
in line \ref{line:POMDP_state_samp}, a belief is chosen randomly from $\hat{\mathcal{B}}_{I}$;
in lines \ref{line:for_a}--\ref{line:hyp_a_2}, the hyperplane associated with each action $a\in \mathcal A$ is computed;
in particular, line \ref{line:hyp_a_1} computes the hyperplane associated with the future value function $V_n(\mathbb B_I(y,a,\beta){,\mathbb I(a,I)})$, for each possible observation $y$ resulting in the belief update $\mathbb B_I(y,a,\beta)$;
 line \ref{line:hyp_a_2} instead performs  the backup operation to determine the new hyperplane of $V_{n+1}(\beta,I)$ associated to action $a$;
 line 8 determines the optimal action that maximizes the value function, 
 so that lines \ref{line:for_a}-\ref{line:val_func} overall approximate the value iteration update $V_{n+1}(\beta,I)=\max_a\mathbb E_{U,Y|a,\beta,I}[\mathcal L_\lambda(U,I,a){+}V_n(\mathbb B_I(Y,a,\beta),\mathbb I(a,I))]$;
in lines \ref{line:hyp_up}-\ref{line:keep_prev}, the new hyperplane and the associated action is added to the set $\mathcal Q_{n+1}^{(I)}$, but only if it yields an improvement in the value function $V_{n+1}(\beta,I){>}\tilde V_{n}(\beta,I)$; otherwise, the previous hyperplane is used;
finally, lines \ref{line:val_func_up}-\ref{line:unimproved} update the set  of un-improved POMDP states based on the newly added hyperplane; only the belief points that have not been improved are part of the next iterations of the algorithm,
and the process continues until the set $\hat{\mathcal{B}}_{I}$ is empty.
Overall, the algorithm guarantees monotonic improvements of the value function in 
$\tilde{\mathcal{B}}_I$. Note that PERSEUS can be executed in parallel by each serving BS, thereby reducing the computation time.
\begin{algorithm}[t]
\DontPrintSemicolon
\SetNoFillComment
\SetKwFunction{Union}{Union}\SetKwFunction{FindCompress}{FindCompress}
\SetKwInOut{Input}{input}\SetKwInOut{Output}{output}
\caption{function {\rm PERSEUS}
\label{alg:alg_1}}
\Input {$I$, $\tilde{\mathcal{B}}_I$, $\{\mathcal Q_n^{(i)}\}_{i\in\mathcal I}$, $
\{a_{\alpha_i}^n,\alpha_i\in\mathcal Q_n^{(i)}\},\forall i \in \mathcal I$, $\lambda$}
	\textbf{Init:} $\tilde V_{n+1}(\beta,I){=}-\infty,\forall \beta\in \tilde{\mathcal B}_I;$  $\hat{\mathcal{B}}_{I}\equiv\tilde{\mathcal{B}}_I;$ $\mathcal Q_{n+1}^{(I)}=\emptyset$\;
	 $\tilde V_{n}(\beta,I)\leftarrow\max_{\alpha_I\in\mathcal Q_n^{(I)}} \inprod{\beta}{\alpha_I^{(r)} - \lambda \alpha_I^{(e)}  }$, and maximizer $(\alpha_{\beta,I}^{(r)},\alpha_{\beta,I}^{(e)}), \forall \beta\in\tilde{\mathcal{B}}_I$\;
	 \While(\tcp*[f]{Unimproved beliefs}){$\hat{\mathcal{B}}_{I} \neq \emptyset$}{
		 Sample $\beta$ from $\hat{\mathcal{B}}_{I}$  (e.g., uniformly)\label{line:POMDP_state_samp}\;
		 \For{each action $a$\label{line:for_a}}{
$\!\!\! I'{=}\mathbb I(a,I)$;
\label{line:hyp_a_1}\!\!\!\!\!\mbox{$\alpha_{y,a}^*{=}\arg\max\limits_{\alpha\in\mathcal Q_n} \inprod{\mathbb B_I(y{,}a{,}\beta)}{\alpha_{I'}^{(r)}{-}\lambda\alpha_{I'}^{(e)}}, \forall y{\in}\mathcal Y$}\;
		$\!\!\!\!\!{\hat\alpha_{a}^*{=}(r(\cdot,I,a),e(\cdot,I,a))}$ $\quad+
  \sum\limits_{u',y}\mathbb P(u',y|\cdot,I,a)
  (\alpha_{y,a}^{*(r)}(u'),\alpha_{y,a}^{*(e)}(u'))$ \label{line:hyp_a_2}}
		Solve $V_{n+1}(\beta,I)=\max_{a\in\mathcal A}
		\inprod{\beta}{\hat\alpha_{a}^{*(r)}-\lambda\hat\alpha_{a}^{*(e)}}$ and maximizing action $a^*$ and $\hat\alpha=\hat\alpha_{a^*}^{*}$ \label{line:val_func}\;
		\eIf(\tcp*[f]{$\hat\alpha$ improves value \label{line:hyp_up}}){$V_{n+1}(\beta,I)>\tilde V_{n}(\beta,I)$}{ 
		$\mathcal Q_{n+1}^{(I)}\leftarrow\mathcal Q_{n+1}^{(I)}\cup\{\hat\alpha\}$; $a_{\hat\alpha}^{n+1}=a^*$ \tcp*[f]{add $\hat\alpha$ to $\mathcal Q_{n+1}^{(I)}$ and define action associated with $\hat\alpha$};\;}
			(\tcp*[f]{keep previous hyperplane $\alpha_{\beta,I}$}){
			
			 $\hat\alpha=\alpha_{\beta,I}$; $\mathcal Q_{n+1}^{(I)}\leftarrow\mathcal Q_{n+1}^{(I)}\cup\{\hat\alpha\}$; $a_{\hat\alpha}^{n+1}=a_{\hat\alpha}^n$\label{line:keep_prev}\;
		}	
			 $\tilde V_{n+1}(\tilde \beta{,} I){\leftarrow}\max\{\inprod{\tilde \beta}{\hat\alpha^{(r)}{-}\lambda\hat\alpha^{(e)}}{,}\tilde V_{n+1}(\tilde \beta{,}I)\}{,}\forall \tilde\beta{\in}\tilde{\mathcal{B}}_I$ \label{line:val_func_up} \;
	  $\hat{\mathcal{B}}_{I} {\leftarrow}
	\{\tilde \beta{\in} \hat{\mathcal{B}}_{{I}} {:} \tilde V_{n+1}(\tilde \beta,I) {<} \tilde V_n(\tilde \beta, I) \}$\tcp*[f]{New set of unimproved beliefs}\label{line:unimproved} \;
	}	
 \Return $\mathcal Q_{n+1}^{(I)}$, 
	$\{a_\alpha^{n+1},\forall\alpha\in\mathcal Q_{n+1}^{(I)}\}$\tcp*[f]{new hyperplanes and associated actions}\; 
	\end{algorithm}


The basic routine for C-PBVI is given in Algorithm~\ref{alg:PBVI}.
However, differently from \cite{DBLP:journals/corr/abs-1109-2145}, we also embed the dual optimization \eqref{dual} by updating the dual variable $\lambda $ in line 6.
In line \ref{line:PERCALL}, we perform one backup operation via PERSEUS (Algorithm \ref{alg:alg_1});
in line \ref{line:VUP}, we compute the new value function $V_{n+1}(\beta,I)$ (based on the new hyperplane sets $\mathcal Q_{n+1}^{(I)}$);
 in line \ref{line:cost}, we compute the approximate cost $\bar E_{n+1}$ starting from state $(\beta_0^*,I_0^*)$,
based on the optimal hyperplane $\alpha^*$;
this is used 
in line \ref{line:lambdaUp} to update the dual variable $\lambda$ via  projected subgradient descent, with the goal to solve the dual problem \eqref{dual} (note that $E_{\max}-\bar E_{n+1}$ is a subgradient of the dual function, see \cite{boyd}):
as a result, $\lambda_n$ is decreased 
if the estimated cost $\bar E_{n+1}<E_{\max}$, to promote throughput maximization over energy cost minimization, otherwise it is increased;
the algorithm continues until the KKT conditions are approximately satisfied \cite{boyd}, i.e.,
$\max_{I\in \mathcal I}\max_{\beta \in \tilde{\mathcal B}_I}|V_{n+1}(\beta,I)-V_{n}(\beta,I)|<\epsilon_V$ (i.e., an approximately fixed point of $V_{n+1}=\tilde H[V_n]$ has been determined and PERSEUS converged),
 $\bar E_{n+1}\leq E_{\max}$ (primal feasibility constraint satisfied) and 
		$\lambda_n|\bar E_{n+1}-E_{\max}|<\epsilon_{E}$ (complementary slackness; note that dual feasibility $\lambda_n\geq 0$ is enforced automatically in line \ref{line:lambdaUp}).
		

After returning the sets of hyperplanes $\{\mathcal Q_{n+1}^{(I)}\}_{I\in\mathcal I}$, the associated actions $\{a_\alpha^{n+1},\forall\alpha\in\mathcal Q_{n+1}^{(I)}\}$, and the dual variable $\lambda_n$, the (approximately) optimal action
to be selected
when operating under the state $(\beta,I)$
 can be computed as
 $$
 \pi^*(\beta,I)=a_{\alpha^*}^{n+1},\ \text{where }\alpha^*=
 	\arg\max\limits_{\alpha\in\mathcal Q_{n+1}^{(I)}} \inprod{\beta}{\alpha^{(r)}-\lambda_n\alpha^{(e)}},
 $$
 along with the approximate expected reward and cost via \eqref{costrew}.
\begin{algorithm}[t]
\DontPrintSemicolon
\caption{Constrained point based value iteration (C-PBVI)}
\label{alg:PBVI}
\textbf{Init:} beliefs $\{ \tilde{\mathcal B}_i \}_{i\in\mathcal I}$; hyperplanes
 $\mathcal Q_0^{(I)}=\{(\mathbf 0,\mathbf 0)\} , \forall I \in \mathcal I$; optimal actions $a_{(\mathbf 0,\mathbf 0)}^0=\mathrm{HO}$;
 value function $V_{n+1}(\beta,i){=}0,\forall \beta {\in}\tilde{\mathcal B}_i, \forall i \in \mathcal I$;
  $\lambda_0\geq 0$; stepsize $\{\gamma_n=\gamma_0/(n+1),n\geq 0\}$\;
	\For{$n=0,\dots$}{
		\For{each $I \in\mathcal I$}{ 
		 $
		(\mathcal Q_{n+1}^{(I)},\{a_\alpha^{n+1},\forall\alpha\in\mathcal Q_{n+1}^{(I)} \})
		= \text{PERSEUS}
		(I,\tilde{\mathcal B}_I,\{\mathcal Q_n^{(I)}\}_{I\in\mathcal I}, \{a_\alpha^n,\alpha{\in}\mathcal Q_n^{(I)}\},\lambda_n)$\label{line:PERCALL}\;
		$V_{n+1}(\beta,I){=}
	\max\limits_{\alpha\in\mathcal Q_{n+1}^{(I)}} \inprod{\beta}{\alpha^{(r)}-\lambda_n\alpha^{(e)}},\forall \beta\in\tilde{\mathcal{B}}_I$\label{line:VUP} \;
	}
		 Let $\bar E_{n+1}=\inprod{\beta_0^*}{\alpha_{\beta_0,I_0}^{(e)*}}$,
		where $\alpha_{\beta_0,I_0}^*{=}
	\arg\max\limits_{\alpha\in\mathcal Q_{n+1}^{(I_0^*)}}\inprod{\beta_0^*}{\alpha^{(r)}-\lambda_n\alpha^{(e)}}$ \label{line:cost}\;
	 $\lambda_{n+1}=\max\{\lambda_n+\gamma_n(\bar E_{n+1}-E_{\max}),0\}$\label{line:lambdaUp}\;
		\If{$\max_{I\in\mathcal I}\max_{\beta \in \tilde{\mathcal B}_I}|V_{n+1}(\beta,I)-V_{n}(\beta,I)|<\epsilon_V$, $\bar E_{n+1}\leq E_{\max}$ and 
		$\lambda_n|\bar E_{n+1}-E_{\max}|<\epsilon_{E}$
		}{ \label{line:checkconv}
			 \Return{$\{\mathcal Q_{n+1}^{(I)}\}_{I\in\mathcal I}$, $\{a_\alpha^{n+1},\forall\alpha\in\mathcal Q_{n+1}^{(I)}\}$, $\lambda_n$}
		}
	}
\end{algorithm}

\begin{figure}[t]
	\centering
	\includegraphics[trim=20 10 20 20, clip,width=0.9\columnwidth]{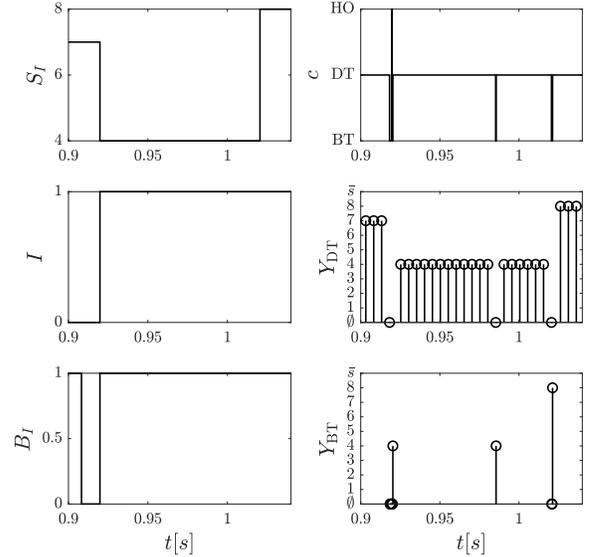}
	\caption{Execution of policy $\pi^*$.} 
	\label{figure:episode}
\end{figure}

In Fig.~\ref{figure:episode}, we plot a time-series of
the following variables for 
a portion of an episode executed under the C-PBVI policy (Algorithms \ref{alg:alg_1} and \ref{alg:PBVI}) under the numerical setup of \secref{sec:numres}, with simulation parameters listed in Table \ref{table1}: serving BS index $I_k$,
BPI $S_k^{(I_k)}$ and blockage state $B_k^{(I_k)}$ of the serving BS $I_k$, the action class $c{\in}\{\rm{DT},\rm{BT},\rm{HO}\}$,
the BT and DT feedbacks $Y_{\rm BT}$ and $Y_{\rm DT}$ as defined in \eqref{btfb} and \eqref{fbdt}. It can be observed in the figure that, at 0.915s, 0.985s and 1.025s, NACKs ($Y_{\rm DT}=\emptyset$) are received after executing the DT action. After each one of these NACKs, the policy executes the BT action. If the BT feedback $ Y_{\rm BT} {\neq} \emptyset$, then DT is performed; otherwise, blockage is detected and  the  HO action is executed. 

It should be noted that, although Algorithm~\ref{alg:PBVI} returns an approximately optimal design, it incurs substantial computational cost in POMDPs with large state and action spaces (hence large number of representative belief points). To remedy this, in the subsequent section we propose simple heuristic policies, inspired by the behavior of the C-PBVI policy described earlier and depicted in Fig. \ref{figure:episode}. These policies will be shown numerically to
trade complexity with sub-optimality and
 achieve satisfactory performance.
 
\vspace{-2mm}
\section{Heuristic Policies}
\label{sec:heuristic} 
In this section, we present two heuristic policies, namely
a belief-based heuristic (B-HEU) and a
finite-state-machine (FSM)-based heuristic (FSM-HEU) and present closed-form expressions for the performance of FSM-HEU. 
Similarly to C-PBVI, B-HEU needs to track the belief $\beta$, whereas 
FSM-HEU is solely based on the current observation signal that defines transitions in a FSM.
For this reason, FSM-HEU has lower complexity than B-HEU, while achieving only a small degradation in performance (see \secref{sec:numres}).
\vspace{-3mm}
\subsection{FSM-based Heuristic policy (FSM-HEU)}
\begin{figure}	
	\centering
	\includegraphics[trim=0 0 0 0, clip,width=0.8\columnwidth]{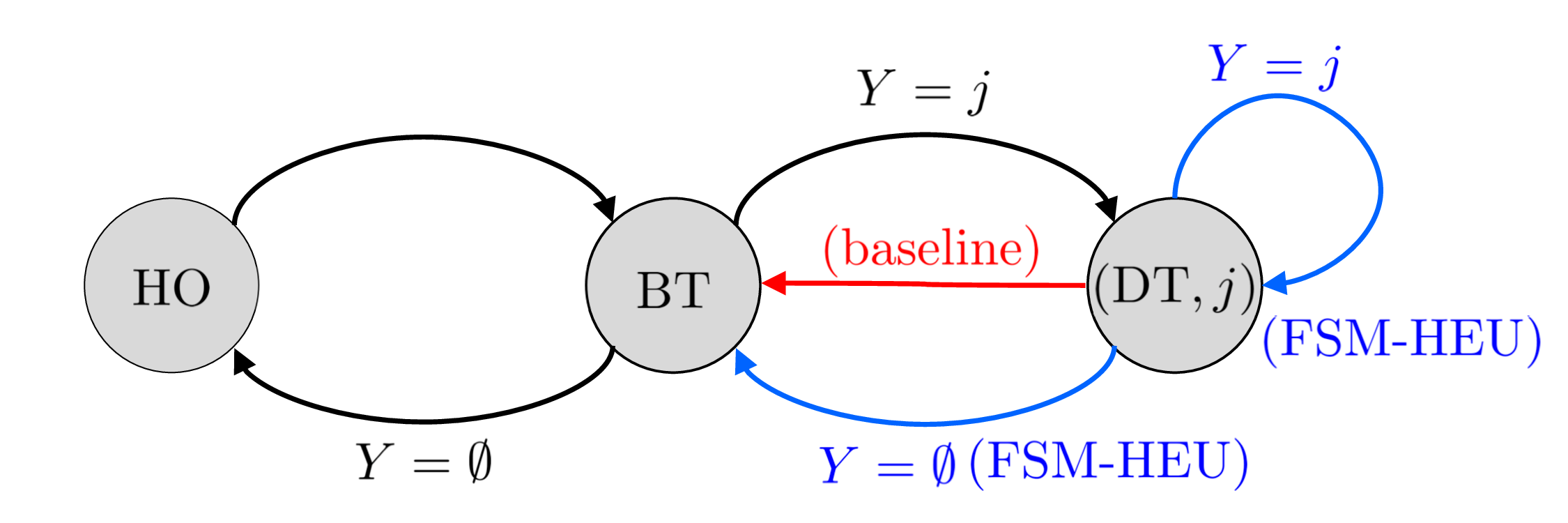}
	\caption{Evolution of the selected action $A_{k}$ of the serving BS based on the observation signal $Y_{k+T}$. Black lines represent the transitions under both FSM-HEU and baseline policies; blue lines represent transitions under the FSM-HEU policy only; the red line represents the transition under the baseline policy only.
		}\label{figure:FSM_HEU}
\end{figure}

The key idea of FSM-HEU is that it selects  actions based solely on a FSM,
whose states define the action to be selected, and whose transitions are defined by the observation signal, as    depicted in Fig.~\ref{figure:FSM_HEU} and described next.
In FSM-HEU, we consider the following actions:
\begin{itemize}[leftmargin=*]
\item the HO action $A_k=(\mathrm{HO},\emptyset, 0, T_{\mathrm{HO}})$ of duration  $T_{\mathrm{HO}}$;
\item the BT action $A_k=(\mathrm{BT},\mathcal S_{I}, \mathrm{SNR}_{\mathrm{BT}}, T_{\mathrm{BT}})$ of duration  $T_{\mathrm{BT}}=|\mathcal S_I|+1$;
in other words, the serving BS performs an exhaustive search over the entire set of SBPIs,
with a fixed SNR $\mathrm{SNR}_{\mathrm{BT}}$ (determined offline), followed by feedback;
\item the $|\mathcal S_I|$ DT actions $(\mathrm{DT},j, \mathrm{SNR}_{\mathrm{DT}},T_{\mathrm{DT}})$, where $j\in\mathcal S_I$;
in other words, the serving BS performs DT 
with  fixed SNR $\mathrm{SNR}_{\mathrm{DT}}$ and duration  $T_{\mathrm{DT}}$ (both determined offline).
\end{itemize}
For notational convenience, we compactly refer to these actions as
$\mathrm{HO}$, $\mathrm{BT}$ and $(\mathrm{DT},j),j\in\mathcal S_I$, respectively.
Let  $A_{k}\in\{{\mathrm{BT}},{\mathrm{HO}}\} \cup \{(\mathrm{DT},j): j\in \mathcal S_I \}$ be the 
selected action of the serving BS $I$ (the state of the FSM at time $k$), of duration $T$, and $Y_{k+T}$ be the observation signal generated by such action, as described in \secref{sec:state_obs_prob};
then, the FSM moves to state $A_{k+T}=\mathbb A_{I}(A_{k},Y_{k+T})$,
which defines the next action $A_{k+T}$ to be selected in the next decision round.
Note that $\mathbb A_I$ defines transitions in the FSM, and the process continues until the episode terminates.

Let us consider the transitions in the FSM, defined by the function $\mathbb A_I$, depicted in Fig.~\ref{figure:FSM_HEU}.
If $A_{k}{=}\mathrm{BT}$ and the observation signal is $Y_{k+T} {=} j{\in}\mathcal S_I$, then the BS  detects the strongest beam $j$; hence FSM-HEU switches to DT  and uses the DT action $A_{k+T}{=}(\mathrm{DT},j){=}\mathbb A_I(\mathrm{BT},j)$ of serving BS $I$ in the next decision round, of duration $T_{\mathrm{DT}}$. On the other hand, if the observation signal is $Y_{k+T}{=}\emptyset$, the BS detects blockage and performs HO to the non-serving BS, so that the new action is $A_{k+T}{=}\mathrm{HO}{=}\mathbb A(\mathrm{BT},\emptyset)$ of serving BS $I$.

If $A_{k}{=}(\mathrm{DT},j)$ of serving BS $I$, i.e., the DT action is executed on beam $j$, of duration $T_{\mathrm{DT}}$, and the signal $Y_{k+T}{=}j$ is observed, then the BS infers that the signal is still sufficiently strong to continue DT on the same beam, and the same action $A_{k+T}{=}(\mathrm{DT},j){=}\mathbb A_I((\mathrm{DT},j),j)$ of the serving BS $I$ is selected again. Otherwise ($Y_{k+T}{=}\emptyset$), the BS detects a loss of alignment, hence the BT action $A_{k+T}{=}\mathrm{BT}{=}\mathbb A_I((\mathrm{DT},j),\emptyset)$ of the serving BS $I$ is executed next.

Finally, if $A_{k}{=}\mathrm{HO}$ of serving BS $I$ (the HO action is chosen, with observation signal $Y_{k+T}{=}\emptyset$), then the new serving BS $I'=1-I$ executes the BT action $A_{k+T}{=}\mathrm{BT}{=}\mathbb A_I(\mathrm{HO},\emptyset)$ next. This procedure continues until the episode terminates.

The performance of FSM-HEU can be computed in closed form. In fact, 
$G_k{=}(U_k,I_k,A_k)$, i.e., the system state $(U_k,I_k)$ and action $A_k$, form a Markov chain, taking values from the state space
\begin{align}
\mathcal G\equiv \bigcup\limits_{I\in\mathcal I}
 \mathcal U\times\{I\}\times[\{{\mathrm{BT}},{\mathrm{HO}}\} \cup \{(\mathrm{DT},j): j\in \mathcal S_I \}].
\end{align}
 To see this, note that
the observation $Y_{k+T}$ and next state $(U_{k+T},I_{k+T})$ (where $T$ is the duration of the selected action $A_k$) have joint distribution given by \eqref{eq:UIsep}, which solely depends on $G_k$; then, in view of the FSM of Fig. \ref{figure:FSM_HEU}, $A_{k+T}=\mathbb A(A_k,Y_{k+T})$ is 
a deterministic function of $A_k$ and $Y_{k+T}$. The state transition probability is
then obtained by computing the marginal with respect to the observation signal $Y_{k+T}$, yielding
\begin{align*}
\label{eq:Pvv}
&\mathbb  P\left(G_{k+T}'=(u',I',a')|G_k=(u,I,a)\right)\\ \nonumber
& {=}\!\!\!\!\!\!\sum_{y \in  \mathcal Y: \mathbb A_I(a,y)=a'}\Bigl[\mathbb P \left(U_{k{+}T}{=}u',Y_{k+T}{=}y|U_k{=}u,I_k=I,A_k{=}a\right)\\
&\qquad\qquad\qquad\times \mathbb P(I_{k+T}=I'|I_k=I,A_k{=}a)\Bigr].
\\
& {=}\sum_{y \in  \mathcal Y}
\mathbb P(u',y |u,I,a)\chi(I'=\mathbb I(a,I))\chi(a'=\mathbb A_I(a,y)).\numberthis
\end{align*}
 We remind that $\mathbb P(u',y |u,I,a)$  is given by \eqref{eq:uoprob_ho}-\eqref{eq:uoprob_dt}.
     Let $\bar R_{\rm{tot}}^{\text{FSM}}(g)$ and $\bar E_{\rm{tot}}^{\text{FSM}}(g)$ be the total expected 
     number of bits delivered and energy cost  under 
     FSM-HEU, starting from state $g$.
 Then, with $\mathbb P(g'|g)$ defined in \eqref{eq:Pvv} and $g=(u,I,a)$,
\begin{align*}
&\bar R_{\rm{tot}}^{\text{FSM}}(u{,}I{,}a) {=}r(u{,}I{,}a ){+}\!\!\!\!\!\!\!\sum_{(u',I',a')\in\mathcal G}\!\!\!\!\!\!\!\mathbb P (u'{,}I'{,}a'|u{,}I{,}a) \bar R_{\rm{tot}}^{\text{FSM}}(u'{,}I'{,}a'),\\
&\bar E_{\rm{tot}}^{\text{FSM}}(u{,}I{,}a) {=}e(u{,}I{,}a ){+}\!\!\!\!\!\!\!\sum_{(u',I',a')\in\mathcal G}\!\!\!\!\!\!\!\mathbb P (u'{,}I'{,}a'|u{,}I{,}a) \bar E_{\rm{tot}}^{\text{FSM}}(u'{,}I'{,}a'),
\end{align*}
where
 $r(\cdot)$ and $e(\cdot)$ are given by \eqref{eq:reward}-\eqref{eq:cost}. 
We can solve these equations in closed form, yielding
\begin{align}
\bar {\mathbf R}_{\rm{tot}}^{\text{FSM}} =(\mathbf I-\mathbf P^{\text{FSM}})^{-1}\mathbf r,\ 
\bar {\mathbf E}_{\rm{tot}}^{\text{FSM}} =(\mathbf I-\mathbf P^{\text{FSM}})^{-1}\mathbf e,
\end{align}
 where $\bar {\mathbf R}_{\rm{tot}}^{\text{FSM}}{=}[\bar R_{\rm{tot}}^{\text{FSM}}(g)]_{g\in\mathcal G}$, $\bar {\mathbf E}_{\rm{tot}}^{\text{FSM}}{=}[\bar E_{\rm{tot}}^{\text{FSM}}(g)]_{g\in\mathcal G}$, $\mathbf r{=}[r(g)]_{g\in\mathcal G}$,
 $\mathbf e{=}[e(g)]_{g\in\mathcal G}$,
 $[\mathbf P^{\text{FSM}}]_{g,g'}{=}\mathbb P(g'|g)$.
\vspace{-3mm}
\subsection{Belief-based Heuristic policy (B-HEU)}
\label{heur:bheu}
\begin{figure}	
	\centering
	\includegraphics[trim=70 60 30 60, clip,width=0.85\columnwidth]{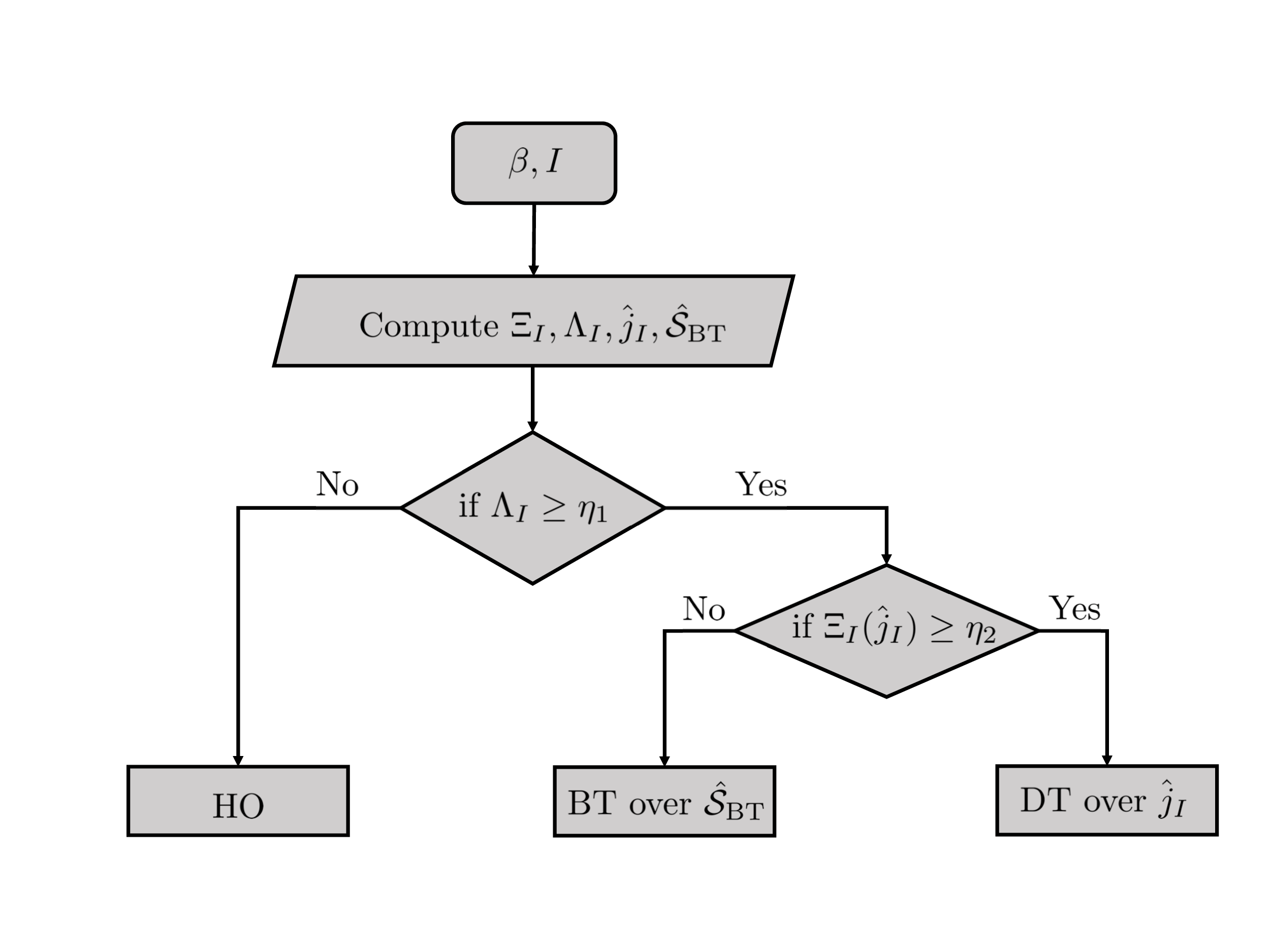}
	\caption{Flow chart for B-HEU Policy.}\label{figure:BHEU_HEU}
\end{figure}
Unlike FSM-HEU, this policy exploits the POMDP state $(\beta_k,I_k)$ in the decision-making process. However, B-HEU selects 
actions in a heuristic fashion as described next, as opposed to C-PBVI (Algorithm \ref{alg:alg_1}), which selects actions (approximately) optimally. The decision making under B-HEU are depicted in the flow chart of Fig.~\ref{figure:BHEU_HEU}.
To describe this policy, let $(\beta,I)$ be the current POMDP state.
Let $\Upxi_I(j)$
be the marginal probability of the UE occupying the $j$th BPI with no blockage under the serving BS $I$, defined as
\begin{align}
\Upxi_I(j)\triangleq \frac{\sum\limits_{(s,b):(s_I,b_I) = (j,1)} \beta(s,b)}{\sum\limits_{j'\in \mathcal S_I}\sum\limits_{(s,b):(s_I,b_I) = (j',1)} \beta(s,b)}.
\end{align}
Then, $\Lambda_I\triangleq\sum\limits_{j\in \mathcal S_I}\sum\limits_{(s,b):(s_I,b_I) = (j,1)} \beta(s,b)$ can be interpreted as the probability of no blockage under the serving BS $I$.
Given these quantities, B-HEU operates as follows, with thresholds $\eta_1$, $\eta_2$ and $\eta_3$ determined offline:
if $\Lambda_I<\eta_1$, then blockage is detected, hence the HO action is selected;
otherwise ($\Lambda_I\geq \eta_1$), let $\hat j_I=\arg\max_{j\in \mathcal S_I}\Upxi_I(j)$ be the most likely BPI occupied by the UE:
if ${\Upxi_I(\hat j_I)}\geq \eta_2$, i.e., the serving BS $I$ is confident that
 the UE belongs to BPI $\hat j_I \in \mathcal S_I$ and there is no blockage, then
the BS performs DT over BPI $\hat j_I $, with SNR $\mathrm{SNR}_{\mathrm{DT}}$ and duration $T_{\mathrm{DT}}$  determined offline.
Otherwise ($\Lambda_I\geq \eta_1$ and ${\Upxi_I(\hat j_I)}< \eta_2$), the BS is uncertain on the BPI of the UE, hence it performs BT over  the smallest BPI set $\hat {\mathcal S}_{\rm BT}$ with aggregate probability greater or equal to $\eta_3$, defined as
\begin{align}
&\hat {\mathcal S}_{\rm BT} \triangleq \arg \! \min_{{\mathcal S}\subseteq \mathcal S_I} |{\mathcal S}|
\ \text{s.t.:}\ \sum_{j\in {\mathcal S}}\Upxi_I
(j)\geq \eta_3.
\end{align}
By doing so, it neglects the least likely set of beams whose aggregate probability is less than $\eta_3$.

After selecting the appropriate action based on the belief, the next serving BS with index $I' = \mathbb I(a,I)$ collects the observation $Y_{k+T}$ and updates its belief using \eqref{eq:belief_update}. Note that, unlike FSM-HEU which performs an exhaustive search during the BT phase, 
B-HEU exploits the current belief $\beta$ to perform BT only on the most likely beams, and therefore reduces the BT overhead. However, it incurs higher complexity than FSM-HEU, since the belief needs to be tracked.
 \vspace{-3mm}
\section{Numerical results}
\label{sec:numres}
In this section, we perform numerical evaluations of the proposed policies. We compare their performance with a baseline policy, which is the same as FSM-HEU except for one key difference: after executing the DT action, it  executes the BT action irrespective of the binary feedback. In other words, $\mathbb A_I((\mathrm{DT},j),Y){=}\mathrm{BT},\forall Y$. Note that, if no blockage is detected, this baseline mimics the periodic exhaustive search. Its performance can be analyzed in closed form in a similar fashion as for FSM-HEU (see its FSM representation in Fig.~\ref{figure:FSM_HEU}). 

The simulation parameters are listed in Table~\ref{table1}. The BSs and UE are both equipped with uniform planar arrays (deployed in the $yz$-plane) with $M_{\rm tx}^{(I)} = M_{{\rm tx},z}^{(I)}\times M_{{\rm tx},y}^{(I)}$  and $M_{\rm rx} = M_{{\rm rx},z}\times M_{{\rm rx},y}$ antennas, respectively. The BS and UE codebooks are based on array steering vectors, designed to provide coverage to a road segment of length $30$m. For numerical simulation, we adopt a blockage dynamic model independent of the UE location, and with blockage states of the two BSs independent of each other. This models a worst-case scenario, where the blockage states of two BSs are independent and they show no correlation with the current and future UE position. In this case, the blockage transition probability can be expressed as $\mathbf B_{b'|bss'} = \mathbf B_{b_0'|b_0}^{(0)} \mathbf B_{b_1'|b_1}^{(1)}$. The transition probabilities can be expressed in terms of average blockage duration $D_{0}^{(I)}$[s] and steady state blockage probability $\pi_0^{(I)}$ as
\begin{align}
\mathbf B_{01}^{(I)} =\frac{\Delta_t}{D_{0}^{(I)}},\
 \mathbf B_{10}^{(I)} = \frac{\pi_0^{(I)}}{1-\pi_0^{(I)}} \frac{\Delta_t}{D_{0}^{(I)}}.
\end{align}
 Using the throughput and power metrics defined in \eqref{metrics},
  the average spectral efficiency (bps/Hz) under policy $\pi$ is
 expressed as $\bar T^{\pi}/W_{\rm tot}$ .
 We choose the initial BS $I=1$ and the initial belief $\beta_0^*(u){=}\chi(u{=}u_0)$, where $u_0{=}(s_0,b_0)$ with $s_0$ denoting the first pair of BS-UE BPI and $b_0=(1,1)$ denoting absence of blockage with respect to both BSs. 
 
\begin{table}[t]
\footnotesize
\begin{center}
\begin{tabular}{|l|l|l|}
\hline
   Parameter & Symbol &  Value \\ \hline
  Number of BS antennas & $M_{\rm tx}^{(I)}$ & $256 = (32\times 8)$ \\
  Number of UE antennas & $M_{\rm rx}^{(I)}$ & $32 = (8\times 4)$ \\
  Number of BS beam & $|\mathcal C_I|$& $8$ \\
  Number of UE beams & $|\mathcal F|$& $8$ \\
  Slot duration & $\Delta_t$ & $100 \mu {\rm s}$ \\ 
  Distance of BS to Rd center & $D$ & $22$m \\ 
 Lane separation & $\Delta_{\rm lane}$ & $3.5$m\\
  BS height & $h_{\rm BS}$ & $10$m\\
  Bandwidth & $W_{\rm tot}$& $100$MHz \\
   Carrier frequency & $f_c$ & $30$GHz \\
    Noise psd &$N_0$ & $-174$dBm/Hz \\
   Noise figure &$F$ & $10$dB \\
   Sidelobe/mainlobe SNR ratio & $\rho$ & -15dB\\
    Fraction of DT slot for & &\\
    channel estimation & $\kappa$ & $0.01$  \\
    HO delay & $T_{\rm HO}$ & 1 slot \\
     DT duration & $T_{\rm DT}$ & $\{20,30,40,50\}$ slots\\
    Steady state blockage prob. & $\pi_0^{(1)},\pi_0^{(2)}$ & $0.2$\\
    Avg blockage duration & $D_{0}^{(1)},D_{0}^{(2)}$ &$200$ms \\
    UE average speed & $\mu_v$ & $30$m/s\\
    UE speed st. dev. &$\sigma_v$ & 10\\
    UE mobility memory param. & $\gamma$ & 0.2\\
    UE lane change prob. & $q_{1\to 2}=q_{2\to 1}$ & 0.01\\
    Accuracy for Algorithm~\ref{alg:PBVI} & $\epsilon_E,\epsilon_V$ & 0.01\\
    B-HEU thresholds  & $(\eta_1,\eta_2,\eta_3)$ & (0.1,0.8,0.60)\\
    \hline
\end{tabular}
\normalsize
\caption{Simulation parameters.\vspace{-3mm}}
\label{table1}
\end{center}
\end{table}

We define a 2D mobility model for a two lane straight highway with lane separation of $\Delta_{\rm lane} = 3.7$m as depicted in Fig \ref{figure:Fig_scenario}.\footnote{The proposed system model and schemes can be used for multi-lane highway with any arbitrary road shape. }
The UE position along the road ($y$-axis) follows a Gauss-Markov mobility model and it changes lanes on the road with probability $q_{l\to l'}$. The speed $V_k$ and position $X_{y,k}$ of the UE along the road ($y$-axis) follow the dynamics
\begin{align}
\label{GMmodel}
\left\{
\begin{array}{l}
V_k=\gamma V_{k-1}+(1-\gamma)\mu_v + \sigma_v\sqrt{1-\gamma^2} \tilde{V}_{k-1},\\
X_{y,k}=X_{y,k-1}+\Delta_{\rm t}V_{k-1},
\end{array}\right.
\end{align}
where, unless otherwise stated, $\mu_v = 30$m/s is the average speed; $\sigma_v = 10$m/s is the standard deviation of speed; $\gamma = 0.2$ is the memory parameter; $\tilde{V}_{k-1}\sim\mathcal N(0,1)$, i.i.d. over slots.
\label{page31}
Note that, under this model, the SBPI $S_k=(s^*_0(X_k),s^*_1(X_k))$ does \emph{not} follow Markovian dynamics, causing a mismatch between the analysis (based on the assumption of Markov state dynamics) and actual state trajectories (which do not follow Markovian dynamics). In addition, there is a mismatch between the sectored antenna model used in the analysis and the actual
beamforming gain, which depends on the beam design and the actual AoA and AoD associated with the current UE position $X_k$ (see \eqref{sigmodel}).
This mismatch might cause the POMDP based policy to underperform. 
 To evaluate the accuracy of our analysis under this more realistic setting,
 in the simulations, we show the results corresponding to the analytical model presented in the paper -- where the transition model $\mathbf S_{s'|s}$ is estimated from simulations of $10,000$ trajectories under the Gauss-Markov model \eqref{GMmodel}, as described in \secref{sec:beam_blockage_dynamics} -- as well as the results obtained through Monte-Carlo simulation using the array steering based analog beamforming and the Gauss-Markov mobility model:
 in this case, the position $X_k$ is generated as in \eqref{GMmodel};
 the beamforming gain is based on the AoA and AoD associated with UE position $X_k$ (see \eqref{sigmodel}) rather than the sectored antenna approximation used in the analytical model (see \secref{sec:sect_antenna});
 the UE's feedback signal $Y_k$ is generated as in \eqref{btfb};
 the belief is then updated using \eqref{eq:belief_update}; actions are selected according to the policy under consideration -- either based on the belief (C-PBVI and B-HEU policies) or feedback signaling (FSM-HEU and baseline policies). 
 Table~\ref{table1} summarizes the numerical parameters.

\begin{figure}[t!]
	\centering
	\includegraphics[trim=10 10 30 10, clip,width=0.9\columnwidth]{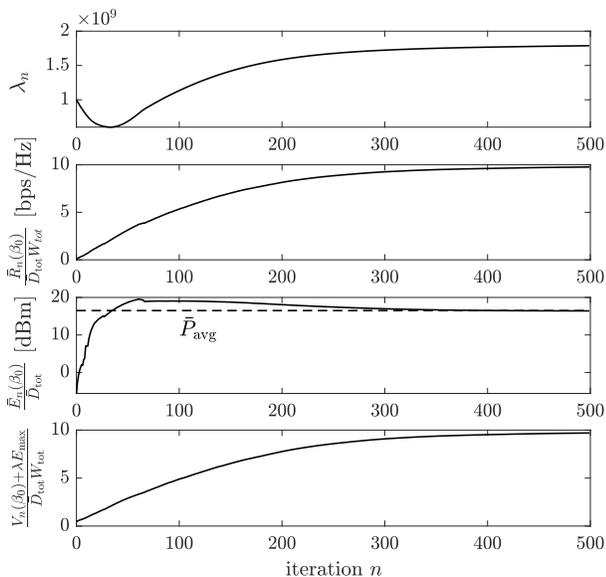}
	\caption{Convergence of C-PBVI Algorithm \ref{alg:PBVI}.}
	\label{figure:conv}
\end{figure}
 \begin{figure}[t]
	\centering
	\includegraphics[trim=10 0 30 15, clip,width=0.9\columnwidth]{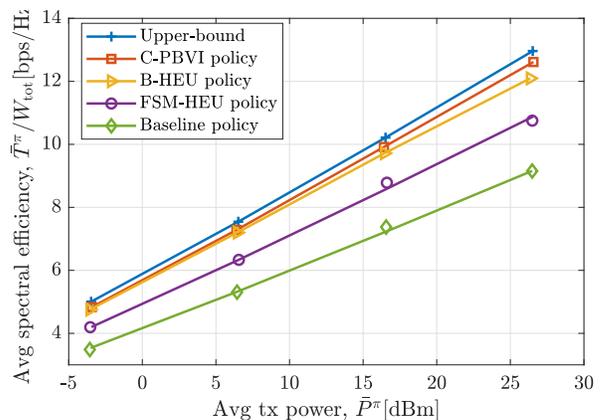}
	\caption{Average spectral efficiency versus average power consumption. The continuous lines represent the analytical curves based on the sectored model and synthetic mobility (generated based on the beam transition probability $\mathbf S_{ss^\prime}$, see Eq.~\eqref{eq:trans_probs_MM}), whereas the markers represent the simulation using analog beamforming and actual mobility.
	} 
	
	\label{figure:SEvP}
\end{figure}  

In Fig.~\ref{figure:conv}, we show the convergence of the \mbox{C-PBVI} Algorithm~\ref{alg:PBVI}, which optimizes both the policy $\pi$ and the dual variable $\lambda$ to meet the power constraint $\bar P^{\pi}\leq\bar P_{\rm avg}$. It can be observed that the dual variable $\lambda$,  expected spectral efficiency $\bar R_{n}/\bar D_{\rm{tot}}/W_{\rm{tot}}$, average power $\bar E_{n}/\bar D_{\rm{tot}}$ and Lagrangian function $[V_n(\beta_0)+\lambda_nE_{\max}]/\bar D_{\rm{tot}}/W_{\rm{tot}}$ converge, and $\bar E_{n}/\bar D_{\rm{tot}}$ converges to the desired average power constraint $\bar P_{\rm avg}=16$dBm.
 In Fig.~\ref{figure:SEvP}, we depict the average spectral efficiency versus the average power consumption. 
 For the heuristic policies, we set $T_{\rm DT}{=}10$ and ${\rm SNR}_{\rm BT}{=}{\rm SNR}_{\rm DT}={\rm SNR}_{\rm pre} M_{\rm tx}^{(I)}M_{\rm rx}, \forall I\in \mathcal I$, where ${\rm SNR}_{\rm pre}$, representing the minimum pre-beamforming SNR, is varied from $-12$dB to $18$dB.\footnote{$M_{\rm tx}^{(I)}M_{\rm rx}$ is the peak beamforming gain for array steering based analog beamforming~\cite{balanis}.}  The upper-bound shown in the figure is obtained by
 a genie-aided policy that always executes DT with perfect knowledge of the state $(u,I)$.  It should be noted that this upper-bound is loose since it is found by assuming perfect state knowledge. The \mbox{C-PBVI} policy $\pi^*$ yields the best performance with negligible performance gap with respect to the upper-bound. It shows a performance gain of up to $4$\%, $17$\% and $38$\% compared to B-HEU, FSM-HEU and baseline, respectively.
 It is also observed that B-HEU shows $12$\% performance gain over  FSM-HEU. On the other hand, the baseline scheme yields up to $24$\% and $15$\% degraded performance compared to B-HEU and FSM-HEU, respectively: in fact, it neglects the DT feedback and instead performs periodic BT, thus incurring significant overhead. We also observe that the curves, obtained through the proposed analytical model, and the markers, representing simulation points obtained considering analog beam design and Gauss Markov mobility, closely match, thereby demonstrating the accuracy of our analysis in  realistic settings.
  
\begin{figure}[t]
	\centering
	\includegraphics[trim=0 0 30 10, clip,width=0.9\columnwidth]{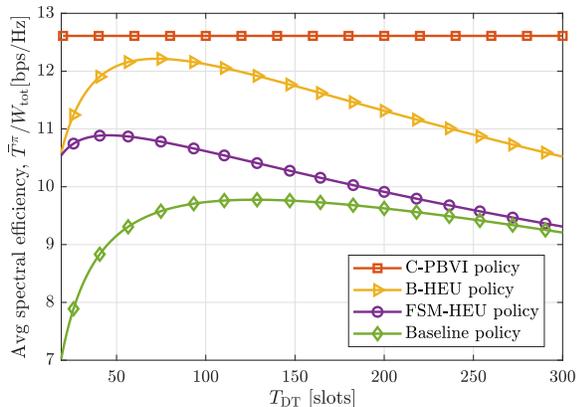}
	\caption{Average spectral efficiency versus $T_{\rm DT}$; ${\rm SNR}_{\rm pre} = 18$dB.} 
	\label{figure:TDT}
\end{figure}

\begin{figure}[t]
	\centering
	\includegraphics[trim=0 0 30 10, clip,width=0.9\columnwidth]{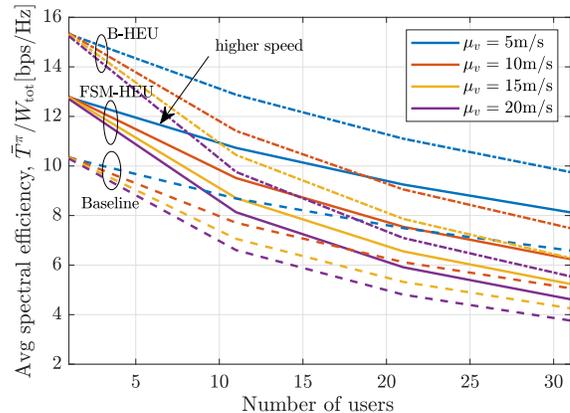}
	\caption{Total average spectral efficiency versus number of UEs for different UE mean speed $\mu_v$; $\sigma_v = 10$m/s, ${\rm SNR}_{\rm pre} = 18$dB, $T_{\rm DT} = 50$.} 
	\label{figure:speed}
\end{figure}

 In Fig.~\ref{figure:TDT}, we plot the spectral efficiency versus the DT time duration $T_{\rm DT}$ used in \mbox{B-HEU}, \mbox{FSM-HEU} and baseline schemes.  As observed previously, the \mbox{C-PBVI} policy outperforms \mbox{B-HEU} and \mbox{FSM-HEU}, and all of them outperform the baseline scheme. \mbox{B-HEU} achieves near-optimal performance with an optimized value of  $T_{\rm DT}{\simeq 70}$[slots] followed by \mbox{FSM-HEU} which performs best  with $T_{\rm DT}{\simeq 40}$. Most remarkably, near-optimal performance is achieved by \mbox{B-HEU} at  a fraction of the complexity of \mbox{C-PBVI}. It is observed that the spectral efficiency initially improves by increasing $T_{\rm DT}$ due to reduced overhead of BT and feedback time. However, after achieving a maximum value at an optimal $T_{\rm DT}$, the spectral efficiency decreases as $T_{\rm DT}$ is further increased. This is attributed to the fact that during very large data transmission periods, loss of alignment and blockages are more likely to occur before the serving BS is able to react to these events. It is also observed that the baseline scheme achieves peak performance at a much higher value of $T_{\rm DT}\simeq 125$[slots].
In fact, since baseline performs periodic BT, it incurs severe overhead, hence there is a stronger incentive to reduce the overhead by extending the duration of DT, as opposed to B-HEU and FSM-HEU which adapt the duration of DT based on the DT feedback signal.
\label{multi-user}
In Fig.~\ref{figure:speed}, we evaluate the impact of mobility and multiple users on blockage dynamics, based on the probabilistic  model developed in \cite{blockage_model}: this model defines a relationship between the dynamics of the blockage process, the number of UEs in the coverage area and their average speed. In fact, mobile UEs may cause time-varying obstructions of the signal (blockages) which may severely degrade the performance of vehicular \mbox{mm-wave} systems, especially in dense and highly-mobile scenarios. In the figure, we plot the total average spectral efficiency versus the number of users and the mean UE speed. The system performance is evaluated via Monte-Carlo simulation. Moreover, we assume that the proposed policies are executed in parallel across multiple UEs, using OFDMA~\cite{oma} to orthogonalize their transmission resources. It can be seen that, for all policies, the spectral efficiency decreases as the mean speed increases: in fact, at higher speed, the UEs not only experience more frequent beam mis-alignments, but also the frequency of occurrence of blockages is exacerbated. The spectral efficiency also degrades as the number of UEs increases: in fact, nearby UEs contribute to creating obstructions and more frequent blockages, as well as a reduced time duration for the unblocked intervals. As previously noted, B-HEU achieves the best performance, followed by FSM-HEU and baseline. Most importantly, the two heuristics B-HEU and FSM-HEU achieve $50$\% and $25$\% higher spectral efficiency than the baseline scheme, respectively, demonstrating their robustness in mobile and dense user scenarios. 

\vspace{-3mm}
\section{Conclusions}
\label{sec:Conclusions}
In this paper, we investigated the design of adaptive beam-training/data-transmission/handover strategies for mm-wave vehicular networks. The mobility and blockage dynamics have been leveraged to obtain the approximately optimal policy via a POMDP formulation and its solution via a constrained point-based value iteration (PBVI) algorithm based on a variation of PERSEUS~\cite{DBLP:journals/corr/abs-1109-2145}. Our numerical results demonstrate superior performance of the C-PBVI policy compared to a baseline scheme with periodic beam-training (up to $38$\% improvement in spectral efficiency). Inspired by the behavior of the \mbox{C-PBVI} policy, we proposed two heuristic policies. These provide low computational alternatives to \mbox{C-PBVI}, with mere performance degradation of $4$\% and $15$\%, and exhibit robustness in scenarios with high density and mobility of users.
\vspace{-3mm}
\section{Acknowledgment}
This work has been supported, in part, by
NSF under grant \mbox{CNS-1642982}, by MIUR (Italian Ministry of Education, University and Research) through the initiative ``Departments of Excellence'' (Law 232/2016) and by the EU MSCA ITN project SCAVENGE ``Sustainable Cellular Networks Harvesting Ambient Energy'' (project no. 675891). The views and opinions expressed in this article are those of the authors and do not necessarily reflect those of the funding institutions.

\vspace{-3mm}
\bibliographystyle{IEEEtran}
\balance
\bibliography{IEEEabrv,bibliography}

\begin{thebibliography}{10}
\providecommand{\url}[1]{#1}
\csname url@samestyle\endcsname
\providecommand{\newblock}{\relax}
\providecommand{\bibinfo}[2]{#2}
\providecommand{\BIBentrySTDinterwordspacing}{\spaceskip=0pt\relax}
\providecommand{\BIBentryALTinterwordstretchfactor}{4}
\providecommand{\BIBentryALTinterwordspacing}{\spaceskip=\fontdimen2\font plus
\BIBentryALTinterwordstretchfactor\fontdimen3\font minus
  \fontdimen4\font\relax}
\providecommand{\BIBforeignlanguage}[2]{{%
\expandafter\ifx\csname l@#1\endcsname\relax
\typeout{** WARNING: IEEEtran.bst: No hyphenation pattern has been}%
\typeout{** loaded for the language `#1'. Using the pattern for}%
\typeout{** the default language instead.}%
\else
\language=\csname l@#1\endcsname
\fi
#2}}
\providecommand{\BIBdecl}{\relax}
\BIBdecl

\bibitem{ICC2020}
M.~{Hussain}, M.~{Scalabrin}, M.~{Rossi}, and N.~{Michelusi}, ``Adaptive
  millimeter-wave communications exploiting mobility and blockage dynamics,''
  in \emph{IEEE International Conference on Communications (ICC)}, 2020, pp.
  1--6.

\bibitem{choi2016millimeter}
J.~Choi, V.~Va, N.~Gonzalez-Prelcic, R.~Daniels, C.~R. Bhat, and R.~W. Heath,
  ``Millimeter-wave vehicular communication to support massive automotive
  sensing,'' \emph{IEEE Communications Magazine}, vol.~54, no.~12, pp.
  160--167, 2016.

\bibitem{Pineau2006PointbasedAF}
J.~Pineau, G.~Gordon, and S.~Thrun, ``Anytime point-based approximations for
  large pomdps,'' \emph{J. Artif. Int. Res.}, vol.~27, no.~1, pp. 335--380,
  Nov. 2006.

\bibitem{DBLP:journals/corr/abs-1109-2145}
M.~T.~J. Spaan and N.~Vlassis, ``Perseus: Randomized point-based value
  iteration for pomdps,'' \emph{J. Artif. Int. Res.}, vol.~24, no.~1, pp.
  195--220, Aug. 2005.

\bibitem{blockage_model}
M.~{Gapeyenko}, A.~{Samuylov}, M.~{Gerasimenko}, D.~{Moltchanov}, S.~{Singh},
  M.~R. {Akdeniz}, E.~{Aryafar}, N.~{Himayat}, S.~{Andreev}, and
  Y.~{Koucheryavy}, ``On the temporal effects of mobile blockers in urban
  millimeter-wave cellular scenarios,'' \emph{IEEE Transactions on Vehicular
  Technology}, vol.~66, no.~11, pp. 10\,124--10\,138, 2017.

\bibitem{michelusi2018optimal}
N.~Michelusi and M.~Hussain, ``Optimal beam-sweeping and communication in
  mobile millimeter-wave networks,'' in \emph{IEEE International Conference on
  Communications (ICC)}, May 2018, pp. 1--6.

\bibitem{marzi}
Z.~Marzi, D.~Ramasamy, and U.~Madhow, ``Compressive channel estimation and
  tracking for large arrays in mm-wave picocells,'' \emph{IEEE Journal of
  Selected Topics in Signal Processing}, vol.~10, no.~3, pp. 514--527, April
  2016.

\bibitem{inverse_finger}
V.~Va, J.~Choi, T.~Shimizu, G.~Bansal, and R.~W. Heath, ``Inverse multipath
  fingerprinting for millimeter wave v2i beam alignment,'' \emph{IEEE
  Transactions on Vehicular Technology}, vol.~67, no.~5, pp. 4042--4058, May
  2018.

\bibitem{7744807}
M.~Giordani, M.~Mezzavilla, and M.~Zorzi, ``Initial access in 5g mmwave
  cellular networks,'' \emph{IEEE Communications Magazine}, vol.~54, no.~11,
  pp. 40--47, November 2016.

\bibitem{va2016beam}
V.~Va, T.~Shimizu, G.~Bansal, and R.~W. Heath, ``Beam design for beam switching
  based millimeter wave vehicle-to-infrastructure communications,'' in
  \emph{IEEE ICC}, 2016, pp. 1--6.

\bibitem{scalabrin2018beam}
M.~{Scalabrin}, N.~{Michelusi}, and M.~{Rossi}, ``Beam training and data
  transmission optimization in millimeter-wave vehicular networks,'' in
  \emph{IEEE Globecom}, Dec 2018, pp. 1--7.

\bibitem{mezzavilla2016mdp}
M.~Mezzavilla, S.~Goyal, S.~Panwar, S.~Rangan, and M.~Zorzi, ``An mdp model for
  optimal handover decisions in mmwave cellular networks,'' in \emph{European
  Conference on Networks and Communications (EuCNC)}.\hskip 1em plus 0.5em
  minus 0.4em\relax IEEE, 2016, pp. 100--105.

\bibitem{pan2012mdp}
J.~Pan and W.~Zhang, ``An mdp-based handover decision algorithm in hierarchical
  lte networks,'' in \emph{IEEE Vehicular Technology Conference (VTC
  Fall)}.\hskip 1em plus 0.5em minus 0.4em\relax IEEE, 2012, pp. 1--5.

\bibitem{stevens2008mdp}
E.~Stevens-Navarro, Y.~Lin, and V.~W. Wong, ``An mdp-based vertical handoff
  decision algorithm for heterogeneous wireless networks,'' \emph{IEEE
  Transactions on Vehicular Technology}, vol.~57, no.~2, pp. 1243--1254, 2008.

\bibitem{alkhateeb2018deep}
A.~{Alkhateeb}, S.~{Alex}, P.~{Varkey}, Y.~{Li}, Q.~{Qu}, and D.~{Tujkovic},
  ``Deep learning coordinated beamforming for highly-mobile millimeter wave
  systems,'' \emph{IEEE Access}, vol.~6, pp. 37\,328--37\,348, 2018.

\bibitem{va2018online}
V.~{Va}, T.~{Shimizu}, G.~{Bansal}, and R.~W. {Heath}, ``Online learning for
  position-aided millimeter wave beam training,'' \emph{IEEE Access}, vol.~7,
  pp. 30\,507--30\,526, 2019.

\bibitem{second-best}
M.~{Hussain} and N.~{Michelusi}, ``Second-best beam-alignment via bayesian
  multi-armed bandits,'' in \emph{IEEE Global Communications Conference
  (GLOBECOM)}, 2019, pp. 1--6.

\bibitem{javdi}
S.~{Chiu}, N.~{Ronquillo}, and T.~{Javidi}, ``Active learning and csi
  acquisition for mmwave initial alignment,'' \emph{IEEE Journal on Selected
  Areas in Communications}, pp. 1--1, 2019.

\bibitem{TWC2019}
M.~{Hussain} and N.~{Michelusi}, ``Energy-efficient interactive beam alignment
  for millimeter-wave networks,'' \emph{IEEE Transactions on Wireless
  Communications}, vol.~18, no.~2, pp. 838--851, Feb 2019.

\bibitem{oma}
M.~{Baghani}, S.~{Parsaeefard}, M.~{Derakhshani}, and W.~{Saad}, ``Dynamic
  non-orthogonal multiple access and orthogonal multiple access in 5g wireless
  networks,'' \emph{IEEE Transactions on Communications}, vol.~67, no.~9, pp.
  6360--6373, 2019.

\bibitem{blockage}
T.~{Bai} and R.~W. {Heath}, ``Coverage analysis for millimeter wave cellular
  networks with blockage effects,'' in \emph{IEEE Global Conference on Signal
  and Information Processing}, Dec 2013, pp. 727--730.

\bibitem{diffuesedTSP}
N.~{Michelusi}, U.~{Mitra}, A.~F. {Molisch}, and M.~{Zorzi}, ``Uwb
  sparse/diffuse channels, part i: Channel models and bayesian estimators,''
  \emph{IEEE Transactions on Signal Processing}, vol.~60, no.~10, pp.
  5307--5319, Oct 2012.

\bibitem{Shim_2014}
D.-S. Shim, C.-K. Yang, J.~Kim, J.~Han, and Y.~Cho, ``Application of motion
  sensors for beam-tracking of mobile stations in mmwave communication
  systems,'' \emph{Sensors}, vol.~14, no.~10, pp. 19\,622--19\,638, Oct 2014.

\bibitem{noh}
S.~{Noh}, M.~D. {Zoltowski}, and D.~J. {Love}, ``{Multi-Resolution Codebook and
  Adaptive Beamforming Sequence Design for Millimeter Wave Beam Alignment},''
  \emph{IEEE Transactions on Wireless Communications}, vol.~16, no.~9, pp.
  5689--5701, Sep. 2017.

\bibitem{LittlesTheorem}
J.~D.~C.~Little and S.~Graves, \emph{Little's Law}, 07 2008, pp. 81--100.

\bibitem{boyd}
S.~P. Boyd and L.~Vandenberghe, \emph{{Convex optimization}}.\hskip 1em plus
  0.5em minus 0.4em\relax Cambridge Univ. Pr., 2011.

\bibitem{balanis}
C.~A. Balanis, \emph{Antenna theory: analysis and design}.\hskip 1em plus 0.5em
  minus 0.4em\relax Wiley, 2016.

\end{thebibliography}

\end{document}